\documentclass[eqsecnum,aps,prd,nofootinbib]{revtex4}
\usepackage[dvips]{graphicx}
\usepackage{amsmath,amssymb,graphicx,mathrsfs}
\usepackage{epsfig}

\newcommand{\be}{\begin{equation}}
\newcommand{\ee}{\end{equation}}
\newcommand{\bea}{\begin{eqnarray}}
\newcommand{\eea}{\end{eqnarray}}
\newcommand{\bg}{{\bar g}}
\newcommand{\G}{{\mathcal G}}
\newcommand{\bxi}{{\bar \xi}}
\newcommand{\bnabla}{{\bar \nabla}}

\newcommand{\bq}{\bar{q}}

\newcommand{\tQ}{{\bar q}}
\newcommand{\non}{\nonumber}
\newcommand{\br}{{\bar r}_0}

\def\({\left(} \def\){\right)}

\def\revise#1       {\raisebox{-0em}{\rule{3pt}{1em}}
                   \marginpar{\raisebox{.5em}{\vrule width3pt\
                     \vrule width0pt height 0pt depth0.5em
                  \hbox to 0cm{\hspace{0cm}{%
                     \parbox[t]{4em}{\raggedright\footnotesize{#1}}}\hss}}}}

\begin{document}
\title{The Weak Gravity Conjecture and the Viscosity Bound \\ with Six-Derivative Corrections}

\author{Aaron J. Amsel}

\email{aamsel@ualberta.ca}

\author{Dan Gorbonos}

\email{dan.gorbonos@mail.huji.ac.il}

\affiliation{Theoretical Physics Institute\\ University of Alberta\\
Edmonton, Alberta, Canada T6G 2G7}

\begin{abstract}
The weak gravity conjecture and the shear viscosity to entropy
density bound place constraints on low energy effective field
theories that may help to distinguish which theories can be UV
completed. Recently, there have been suggestions of a possible
correlation between the two constraints. In some interesting cases,
the behavior was precisely such that the conjectures were  mutually
exclusive. Motivated by these works, we study the mass to charge and
shear viscosity to entropy density ratios for charged AdS$_5$ black
branes, which are holographically dual to four-dimensional CFTs at
finite temperature. We study a family of four-derivative and
six-derivative perturbative corrections to these backgrounds. We
identify the region in parameter space where the two constraints are
satisfied and in particular find that the inclusion of the
next-to-leading perturbative correction introduces wider
possibilities for the satisfaction of both constraints.
\end{abstract}

\maketitle


\tableofcontents

\section{Introduction}
\label{intro} The weak gravity conjecture (WGC) \cite{weak1} and the
shear viscosity to entropy density bound (KSS) \cite{KSS} have been
suggested as tests to distinguish theories in the landscape from
theories that belong to the
swampland~\cite{Vafa,blmsy,Cremonini:2009ih}. As argued
in~\cite{Vafa}, any quantum gravity theory imposes certain
constraints on low energy physics, so that not every effective
theory can be UV completed. Thus, effective theories can be
classified as ``good'' theories that admit a valid UV completion
(landscape) or ``bad'' theories that cannot be consistently
completed (swampland).

Even though both conjectures (WGC and KSS) serve to place
constraints on effective theories, there are some important
differences between the two. The WGC was formulated for gravity in
asymptotically flat spacetime, while the KSS bound was formulated
for asymptotically AdS spacetime. The WGC deals with physical,
global quantities that are usually defined at the asymptotic
boundary of the spacetime (mass and charge), while the KSS bound
deals with local quantities which are defined at the horizon
(viscosity and entropy density). Nevertheless, there have been
attempts to extend the applicability of the WGC to asymptotically
AdS backgrounds as
well~\cite{Kats2,Cremonini,MyersChemical,Cremonini:2009ih,Pal}. As
explained further below, these works also made the intriguing
suggestion that these conjectures might in fact be correlated to
each other in some way.  In this work, we shall further explore the
interplay of the two constraints in a toy model of an effective
theory with four-derivative and six-derivative corrections.

\begin{figure}[hbt]
\begin{center}
\includegraphics[width=3.8in]{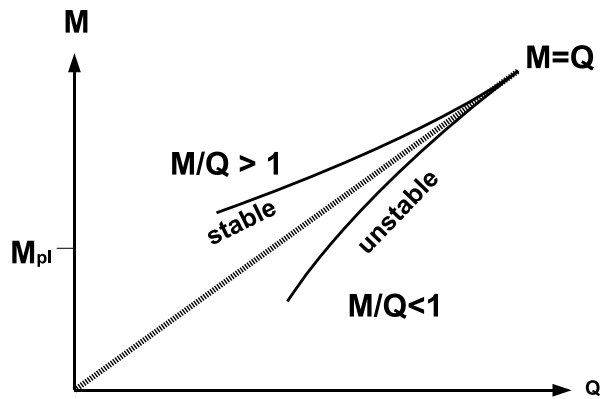}
\end{center}
\caption{The two a-priori possibilities for the quantum corrections
to extremal black holes. The unstable branch is the one which is
expected to occur in consistent theories of quantum gravity.} \label{fig1}
\end{figure}

One implication of the WGC is that higher derivative corrections in
a consistent theory of quantum gravity should reduce the mass to
charge ratio of extremal black holes\footnote{In what follows, the
term ``WGC'' will be meant only to refer to this particular aspect
of the conjecture.}~\cite{weak1,weak2}. We give here a brief version
of the argument for this statement. For classical charged black
holes (e.g. Reissner-Nordstr\"{o}m), the minimal value of the mass
to charge ratio is achieved when the black hole is extremal, while
going below this minimal value creates a naked singularity. We
assume that the existence of a large number of stable black hole
states is unnatural unless there is a global symmetry (e.g.
supersymmetry) that protects them from quantum corrections and/or
decay.  One may then further argue that such a symmetry should not
exist in a consistent theory of quantum gravity~\cite{weak1}.
Therefore, quantum corrections should take the parameters of the
black hole away from their classical extremal values. As an
additional consequence, any black hole (even at extremality) should
be allowed to decay. Let us choose units in which the classical
value of the mass to charge ratio at extremality is 1.  Suppose that
quantum corrected black holes satisfy $M/Q>1$ and that they are
unstable. Then a given black hole will decay into two black holes
with masses $M_1$,$M_2$ and charges $Q_1$, $Q_2$ such that
$M_1+M_2<M$, $Q_1+Q_2=Q$. According to our assumption, the decay
products should also satisfy $M_i/Q_i >1$ ($i=1,2$) (see
Figure~\ref{fig1}, which reproduces a similar figure
from~\cite{weak2}). Since $M_i<M$,  the quantum correction is
larger, which implies $M_i/Q_i>M/Q$. Then it follows that
$M_1+M_2>M$, which contradicts our assumption about the instability
of this branch. Hence, the second branch (where $M/Q<1$) is unstable
and is expected to occur in a consistent quantum gravity theory.

The perturbed heterotic string states were given in~\cite{weak1} as
evidence in favor of the conjecture. These states are non-supersymmetric and approach the line $M=Q$
from below as expected from the WGC. An analysis of the mass to
charge ratio for general four-derivative corrections to
non-supersymmetric black holes in asymptotically flat spacetime was
carried out in~\cite{weak2}. The results were consistent with the
conjecture for those cases in which the values of the four-derivative couplings
are known (four dimensions). Further support for the conjecture was
given in~\cite{banks}, based on semi-classical considerations.

If the WGC is true, it is expected to apply also to non-extremal
black holes close to extremality, since we expect theories to behave smoothly as the parameters vary. The conjecture was examined
for four-dimensional non-extremal black holes with two electric
charges, which are solutions that correspond to fundamental strings
with generic momentum and winding on an internal
circle~\cite{weak3}. The results showed that for this case the mass
to charge ratio is smaller compared to the uncorrected ratio for any value
of the charges where a regular black hole solution exists. These
results were later extended to $d$ dimensions~\cite{weak4} with the same
conclusions.

The KSS bound was presented as a conjecture for field theories that
have a holographic dual in~\cite{KSS}. The conjecture suggests that
the ratio of the coefficient of shear viscosity $\eta$ to entropy
density $s$ has a lower bound and that this lower bound is $1/4\pi$.
The bound is saturated for boundary field theories in the limit of
infinite 't Hooft coupling $\lambda$ and number of colors $N_c$.
Such theories are dual to Einstein gravity (without corrections).
The authors of~\cite{KSS} also gave a general argument that the
ratio should be greater than some constant of order one. A short version of the argument is as follows. The product of the energy $\epsilon$ of
a particle in the fluid and its mean free time
$\tau_{mft}$ is, according to Heisenberg's uncertainty principle,
$\epsilon \tau_{mft}>\hbar$. The viscosity is proportional to $n
\epsilon \tau_{mft}$, where $n$ is the density of particles. The
entropy density is also proportional to $n$, with $s \sim k_{B}n$. Then the
viscosity to entropy density ratio is $\eta / s \sim \hbar/k_{B}$, and since
we take $\hbar=k_{B}=1$, the constant is of order one.

Another piece of evidence supporting the conjecture arose from the
explicit computation of the leading $\alpha'$ correction for type
IIB string theory compactified to five dimensions ($R^4$
corrections)~\cite{buchel,buchel2,buchel3}. This computation showed
that the correction increased the shear viscosity to entropy density
ratio.

However, more general computations in higher derivative gravity
showed that the KSS bound can be violated, although the crucial sign
of the coefficient in front of the higher derivative correction is
in general undetermined. Examples of models where there is a
possible violation of the KSS bound were given
in~\cite{blmsy,Kats2,Cremonini,ssp,bms,sm,ag,Ge1,Cai1,Cai2,Kout,deBoer,
Camanho,Camanho2}. Violations of the bound might be related to
inconsistencies of the boundary theory, for example by introduction
of ghosts~\cite{bm}. Even if violation of the KSS bound occurs in
some models, there may still exist an improved lower bound. A bound
of $4/25\pi$ was proposed in \cite{causality} based on causality
violation in the CFT. Note that this bound is based on the
consistency of the boundary theory and is still quite close to the
original bound. For the purposes of this work, we will usually take
the bound to be $1/4\pi$. Even if the bound were different, as
explained above, it would be of the same order and the qualitative
conclusions often remain the same. We will, however, point out below
some instances where the behavior does change if the bound is
relaxed to $4/25\pi$.

Suggestions of correlation between the two constraints appeared
in~\cite{Kats2,Cremonini,MyersChemical,Cremonini:2009ih, Pal}.
In~\cite{Cremonini:2009ih}, examples with four-derivative
corrections were studied in a five-dimensional asymptotically AdS
spacetime. The examples suggested that the bounds cannot be
satisfied simultaneously. In particular, the  WGC and KSS bound
require opposite signs of the coefficients of certain corrections in
four-derivative gravity.
The precise statement depends of course on the relations (if any)
between the coefficients. In certain cases of interest, Weyl anomaly
matching shows that the coefficients are proportional to the
difference of the two central charges $a,c$ in the dual four
dimensional CFT. As a result, it appeared that the WGC required
$c-a>0$, but this is exactly the condition that implies violation of
the KSS bound. This is true for naturally appearing corrections like
the Gauss-Bonnet term \be R_{GB}^2\equiv
R_{\mu\nu\rho\sigma}R^{\mu\nu\rho\sigma}-4\,R_{\mu\nu}R^{\mu\nu}+R^2
\ee or Weyl-tensor-squared corrections~\cite{Cremonini:2009ih}
(motivated by the general form of supersymmetric higher derivative
actions) \be W^2 \equiv
\frac{1}{6}R^2-\frac{4}{3}R_{\mu\nu}R^{\mu\nu}+R_{\mu\nu\rho\sigma}R^{\mu\nu\rho\sigma}.
\ee There are some special cases in four-derivative gravity in which
the two bounds are satisfied together, such as when one takes the
only correction to be
$R_{\mu\nu\rho\sigma}\,F^{\mu\nu}\,F^{\rho\sigma}$~\cite{Pal}.
However, we will not consider this term in our toy model family (as
discussed below).

It is important to note that in the two examples listed above (and
indeed for generic four-derivative corrections excluding the
$R_{\mu\nu\rho\sigma}\,F^{\mu\nu}\,F^{\rho\sigma}$ term), the
correction to the shear viscosity to entropy density ratio vanishes
in the extremal limit. In such cases, it is natural to consider
$\eta/s$ for non-extremal cases in order to see the direction from
which the bound is approached.  To gain additional perspective on
the apparent tension between the bounds, it is also interesting
though to find further examples in which the shear viscosity to
entropy density ratio does not vanish at extremality. As we will
see, the first order at which this phenomenon occurs is in many
cases at the level of curvature-cubed corrections (see
also~\cite{Dutta2}).

Six-derivative gravity is also important for studying the effects of
higher derivative corrections that capture more parameters of the
dual CFT at finite 't Hooft coupling and number of
colors~\cite{Myers1,Myers2}. For example, curvature-cubed
corrections in the bulk are required to fully characterize the
energy flux one-point function of the CFT~\cite{Hofman:2008ar}.
Furthermore, since curvature-cubed terms break supersymmetry, these
theories are relevant to the study of non-supersymmetric CFTs
\cite{Myers2}.

In general, when we want to consider the WGC and KSS bound for black
holes with parameters closer to the Planck scale, the next
correction after four-derivatives will be six-derivatives. Of
course, all the higher derivative corrections are important for the
full quantum gravity description in the vicinity of the Planck
scale.  We can, however, consider an  intermediate  small region
just beyond the regime where the four-derivative terms dominate. In
this region, the six-derivative corrections become of the same order
as the four-derivative corrections, but higher orders are still
negligible.  Below, we shall study the two constraints in this
region, which is a further step towards the quantum gravity regime.

In order to explore new possibilities for satisfying both
constraints in six-derivative actions, we take a toy model that
includes two curvature-cubed terms which are invariant under field
redefinitions: \be I_1= R^{\mu \nu}{}{}_{\alpha \beta} R^{\alpha
\beta}{}{}_{\lambda \rho} R^{\lambda \rho}{}{}_{\mu \nu}\,, \quad
I_2=  R^{\mu \nu}{}{}_{\rho \sigma} R^{\rho \tau}{}{}_{\lambda \mu}
R^{\sigma}{}_{\tau}{}^\lambda{}_\nu.
 \ee
These are the only terms required to describe curvature-cubed
gravity corrections (up to field redefinitions) when charge is not
included~\cite{tseytlin}.  For simplicity, we do not include higher
derivative corrections involving the Maxwell field in our toy model.
In other words, we add the charge only to the Einstein-Hilbert part
of the action. We leave the general six-derivative action for future
study, as there are many additional terms that involve Maxwell fields.  As we will see, the
two six-derivative terms that we do include already provide a much richer picture than the
restriction to only four-derivative terms.

Explicitly, we take for our toy model an action of the form
\begin{eqnarray}
\label{action} S &=&\frac{1}{16\,\pi\,G}\int
d^5x\sqrt{-g}\left[R-\frac{1}{4}F^2 -2 \Lambda+b_1 R^2 +b_2
R^2_{GB}+b_3\,R_{\mu \nu}\,R^{\mu\nu}+c_1\,I_1+c_2\,I_2 \right].
\end{eqnarray}
 One may view an action of this form as
an effective theory arising in a string theory $\alpha'$ expansion,
with $b_i \sim O(\alpha')$ and $c_i \sim O(\alpha'^2)$. In theories
without charge, we can eliminate the $b_1$ and $b_3$ terms by field
redefinitions and write the action only with $b_2$ (the Gauss-Bonnet
term). When we include charge, however, these terms cannot be
eliminated as the field redefinitions would then generate $R\,F^2$
and $R_{\mu\nu}\,F^{\mu\lambda}\,F^{\nu}_{\phantom{\nu}\lambda}$
terms~\cite{Cremonini:2009ih}.

The paper is organized as follows.  In section~\ref{mass_section},
we compute the perturbative corrections to the mass to charge ratio
by applying the covariant ADT
method~\cite{Abbott:1981ff,Deser:2002rt} to a $d$ dimensional
extension of~\eqref{action}. In section~\ref{viscosity_section}, we
compute the perturbative corrections to the shear viscosity to
entropy density ratio in our toy model. Next, in
section~\ref{analysis}, we analyze the WGC and KSS bound to
determine the influence of the $O(\alpha'^2)$ corrections on various
cases of the action \eqref{action}. In particular, for various toy
models we provide precise plots of regions in the parameter space
where the two bounds are satisfied, i.e. where the theories are
``good.''  Our main conclusion is that six-derivative terms give us
more possibilities to satisfy the constraints and that the apparent
tension observed in four-derivative gravity may disappear when
viewed from the higher order perspective.  We conclude with a brief
discussion of our results and future directions in
section~\ref{conclusions}.

\section{The Mass to Charge Ratio}
\label{mass_section} In this section, we calculate the mass to
charge ratio of a static, asymptotically AdS black hole solution of
a six-derivative gravity theory specified below.  We begin by giving
a brief introduction to a general method (referred to as
``Abbot-Deser-Tekin'' or ``ADT'' method) for calculating the energy
in higher derivative gravity \cite{Deser:2002rt,Deser:2002jk}.  We
then apply this procedure to the particular case of an action with
terms up to curvature-cubed. Finally, we couple the theory to a
gauge field and obtain a perturbative expression for the mass to
charge ratio of a charged black hole solution.


\subsection{The ADT Method}

The celebrated result of ADM \cite{ADM} for energy in
Einstein-Hilbert gravity with asymptotically flat boundary
conditions was generalized to spacetimes with a cosmological
constant in \cite{Abbott:1981ff}. These so-called ``AD charges''
were written in a manifestly covariant way and once again could be
expressed as pure surface integrals.  The method used to construct
the AD charges was then further generalized to arbitrary higher
curvature theories in \cite{Deser:2002rt,Deser:2002jk}.

This ADT method is similar in spirit to the
Landau-Lifshitz pseudotensor method for calculating energy \cite{LL}
in asymptotically flat curved spacetime.  In particular, one
proceeds by linearizing the equations of motion with respect to a
background spacetime (AdS in our case).  This leads to an effective
stress-energy tensor that consists of matter sources and terms
higher order in the perturbation.  This tensor turns out to be
covariantly conserved and can thus be used to construct a conserved
charge associated with an isometry of the background.

As we will see, the ADT method involves relatively little formalism
and is computationally straightforward. In addition, this method has
the advantage of not involving any explicit regularization or
subtraction of infinities, as required in counter-term methods (see
e.g. \cite{Henningson:1998gx,Balasubramanian:1999re}). Unlike
Euclidean path integral techniques (e.g. \cite{Hawking:1980gf}), the
ADT framework naturally gives the gravitational mass as an integral
at asymptotic infinity, without any need to identify a horizon in
the interior.  For perturbations that vanish sufficiently fast at
asymptotic infinity\footnote{Here we mean that the perturbation
about a solution falls off fast enough at infinity that the theory
is asymptotically linear, i.e. that the linearized equations of
motion are obeyed near infinity.  In this case, the charges of the
linearized theory can be used to obtain the charges of the
non-linear theory.  This condition indeed holds for the case of
standard asymptotically AdS boundary conditions
\cite{Henneaux:1985tv} that we consider in this work.}, the ADT
charges are exactly the same as the charges derived using the
covariant phase space methods of
\cite{Barnich:2001jy,Barnich:2007bf,Compere:2007az}, which in turn
differ from the charges of Wald et al.
\cite{Lee:1990nz,wald2,Wald:1999wa} by a surface term proportional
to the Killing equations.

Let us consider some arbitrary
gravitational theory with equations of motion of the form
\begin{equation}
\label{geneom} \Phi_{\mu \nu}(g,R,\nabla R, R^2,\ldots) = \kappa
\tau_{\mu \nu},
\end{equation}
where $\kappa$ is the gravitational coupling and $\tau_{\mu \nu}$ is
the matter stress-energy tensor.  The symmetric tensor $\Phi_{\mu
\nu}$, which is the analogue of the Einstein tensor, may depend on
the metric, the curvature, derivatives of the curvature, and various
combinations thereof.  Assuming that the action is invariant under
diffeomorphisms, we obtain the geometric identity $\nabla^\mu
\Phi_{\mu \nu}=0$ (the generalized Bianchi identity) and the
covariant conservation of the stress tensor $\nabla^\mu \tau_{\mu
\nu}=0$.

Now, we further assume that there exists a background solution
$\bg_{\mu \nu}$ to the equations \eqref{geneom} with $\tau_{\mu
\nu}=0$.  Then we decompose the metric as
\begin{equation}
g_{\mu \nu}=\bg_{\mu \nu}+h_{\mu \nu} \,,
\end{equation}
where we note that the deviation $h_{\mu \nu}$ is not necessarily
infinitesimal, but again is required to fall off sufficiently fast
at infinity.  By expanding the left-hand side of \eqref{geneom} in
$h_{\mu \nu}$, the equations of motion may be expressed as
\begin{equation}
\phi^{(1)}_{\mu \nu}= \kappa \tau_{\mu \nu}-\phi^{(2)}_{\mu
\nu}-\phi^{(3)}_{\mu \nu}\ldots\equiv \kappa T_{\mu \nu}\,,
\end{equation}
where $\phi^{(i)}_{\mu \nu}$ denotes all terms in the expansion of
$\Phi_{\mu \nu}$ involving $i$ powers of $h_{\mu \nu}$ and we have defined the
effective stress-tensor $T_{\mu \nu}$.  It then follows from the
Bianchi identity of the full theory that $\bnabla^\mu
\phi^{(1)}_{\mu \nu}=0=\bnabla^\mu T_{\mu \nu}$.

Suppose that the background spacetime admits a timelike Killing
vector $\bar{\xi}^\mu$ and let $\Sigma$ be a constant-time
hypersurface with unit normal $n^\mu$.  Then we can construct a
conserved energy in the standard way
\begin{equation}
\label{ebulk} E=\int_\Sigma d^{d-1}x\sqrt{\bg_\Sigma}\, n_\mu T^{\mu
\nu}\bar{\xi}_\nu \,,
\end{equation}
where $\bg_\Sigma$ denotes the determinant of the induced metric on
$\Sigma$. Because $\bnabla^\mu (T_{\mu \nu} \bar \xi^\nu)=0$, it
follows that $T_{\mu \nu} \bar \xi^\nu=\bnabla^\nu \mathcal{F}_{\nu
\mu}$ for some antisymmetric tensor $\mathcal{F}_{\nu \mu}$. The
bulk integral \eqref{ebulk} can therefore be rewritten as a surface
integral over the boundary $\partial \Sigma$
\begin{eqnarray}
\label{esurf} E=\int_{\partial\Sigma}
d^{d-2}x\sqrt{\bg_{\partial\Sigma}} \,n_\mu r_\nu \mathcal{F}^{\nu
\mu} \,,
\end{eqnarray}
where $r_\mu$ is the unit normal to the boundary.

In summary, to apply the ADT method, one linearizes the equations of
motion to obtain the stress-energy tensor, and then expresses the
conserved current $T^{\mu \nu} \bar \xi_\nu$ as a total derivative
to find the ``potential'' $\mathcal{F}^{\nu \mu}$.  Note that by
construction, the background spacetime $\bg_{\mu \nu}$ has $E=0$.


\subsection{Energy in Six-Derivative Gravity}
We now wish to apply the ADT procedure to a six-derivative theory
that we describe below.  The case of a generic four-derivative
theory has been worked out in detail previously \cite{Deser:2002jk},
so it remains to apply the method only to the curvature-cubed terms
we wish to add.  The only potentially non-trivial step is to rewrite
the conserved current as a total derivative, but we will see that
there is a simplification below.

Let us consider a theory of the form
\begin{eqnarray}
\label{action2}
S &=&\frac{1}{16\,\pi\,G}\int d^dx\sqrt{-g}\Big[R-2 \Lambda+b_1 R^2 +b_2 (R^2_{\mu \nu \rho \sigma}-4 R^2_{\mu \nu}+R^2) +b_3 R^2_{\mu \nu}  \nonumber \\
&& \qquad \qquad \qquad +c_1 R^{\mu \nu}{}{}_{\alpha \beta}
R^{\alpha \beta}{}{}_{\lambda \rho} R^{\lambda \rho}{}{}_{\mu
\nu}+c_2 R^{\mu \nu}{}{}_{\rho \sigma} R^{\rho \tau}{}{}_{\lambda
\mu} R^{\sigma}{}_{\tau}{}^\lambda{}_\nu \Big] +S_{m}\,,
\end{eqnarray}
where the matter action
\begin{equation}
S_{m}= \frac{1}{16\,\pi\,G}\int d^dx\sqrt{-g} L_m
\end{equation}
is at this point arbitrary. The corresponding
equation of motion is
\begin{equation}
\label{eom} R_{\mu \nu}-\frac{1}{2} g_{\mu \nu} R+\Lambda g_{\mu
\nu}+\Phi^{(4)}_{\mu \nu}+\Phi^{(6)}_{\mu \nu}= -\frac{ 16 \pi G}{\sqrt{-g}}
\frac{\delta S_m}{\delta g^{\mu \nu}} = 16 \pi G \, \tau_{\mu \nu}
\end{equation}
where
\begin{eqnarray}
\Phi^{(4)}_{\mu \nu}&=&2 b_1 R \left(R_{\mu \nu} -\frac{1}{4} R
g_{\mu \nu}\right)+(2 b_1 +b_3)(g_{\mu \nu}\Box -\nabla_\mu
\nabla_\nu)R+ b_3 \Box \left(R_{\mu \nu} -\frac{1}{2} R g_{\mu
\nu}\right)\nonumber\\&& +2b_2\left(R R_{\mu \nu}-2 R_{\mu \sigma
\nu \rho}R^{\sigma \rho}+ R_{\mu \sigma \rho \lambda}R_\nu{}^{\sigma
\rho \lambda}-2 R_{\mu
\sigma}R^{\sigma}{}_\nu-\frac{1}{4}(R^2_{\lambda \kappa \rho
\sigma}-4 R^2_{\rho \sigma}+R^2)g_{\mu \nu}\right)
\nonumber \\
&&+2 b_3\left(R_{\mu \sigma \nu \rho}-\frac{1}{4} R_{\sigma \rho}
g_{\mu \nu}\right)R^{\sigma \rho}
\end{eqnarray}
\begin{eqnarray}
\Phi^{(6)}_{\mu \nu}&=&\frac{1}{2}\Big[-6 c_1\nabla_\kappa
\nabla_\rho (R_{\mu}{}^{\rho \lambda \tau} R_{\lambda
\tau}{}^\kappa{}_\nu )
+3 c_2 \nabla_\kappa \nabla_\rho(R_{\mu \tau \lambda \nu} R^{\rho \tau \lambda \kappa} - R_\mu{}^{\tau \lambda \kappa} R^\rho{}_{\tau \lambda \nu})\nonumber\\
&&+c_1(3 R_{\sigma \mu \lambda \rho}R^{\lambda \rho \alpha \beta}
R^{\sigma}{}_{\nu \alpha \beta}-\frac{1}{2}
R^{\sigma \gamma}{}{}_{\lambda \rho}R^{\lambda \rho \alpha \beta} R_{\alpha \beta \sigma \gamma} g_{\mu \nu})\nonumber \\
&&+c_2(3 R^\beta{}_{\mu \rho \sigma} R^{\rho \tau}{}_{\lambda \beta}
R^{\sigma}{}_\tau{}^\lambda{}_\nu-\frac{1}{2} R^{\beta \gamma}{}_{
\rho \sigma} R^{\rho \tau}{}_{\lambda \beta}
R^{\sigma}{}_\tau{}^\lambda{}_\gamma g_{\mu \nu}\Big]+\frac{1}{2}
\Big[\mu \leftrightarrow \nu\Big] \,.
\end{eqnarray}

We now look for an exact AdS solution of these equations with no
matter fields.  Using that in this case the Riemann tensor takes the
maximally symmetric form
\begin{equation}
\bar R_{\mu \rho \nu \sigma}= \frac{2 \Lambda_{eff}}{(d-1)(d-2)}\,
(\bg_{\mu \nu} \bg_{\rho \sigma}-\bg_{\mu \sigma} \bg_{\rho \nu})\,,
\end{equation}
we find that there can exist an AdS solution with an ``effective''
cosmological constant $\Lambda_{eff}$ satisfying the cubic equation
\begin{eqnarray}
\label{lambda}
0&=&\frac{8}{(d-2)^3(d-1)^2}\left[2(6-d)c_1+\frac{(6-d)(d-2)}{2}c_2\right]\Lambda_{eff}^3 \nonumber\\
&&-2\left[\frac{(d-4)(d b_1 +b_3)}{(d-2)^2}+\frac{(d-3)(d-4)
b_2}{(d-2)(d-1)}\right] \Lambda_{eff}^2 -\Lambda_{eff}+\Lambda\,.
\end{eqnarray}
The perturbative solution of this equation takes the form
$\Lambda_{eff} = \Lambda+\ldots$, where the explicit expressions for
the corrections are given in appendix \ref{lin}.

The next step is to linearize the equations of motion \eqref{eom}
with respect to the background AdS solution by writing $g_{\mu
\nu}=\bg_{\mu \nu}+h_{\mu \nu}$. Using the results of
\cite{Deser:2002jk} and of appendix~\ref{lin}, we find that the
stress tensor is
\begin{eqnarray}
\label{stress} 16\,\pi\,G T_{\mu \nu}= \alpha_1 \G^L_{\mu
\nu}+\alpha_2 \left(\bg_{\mu \nu} \bar{\Box} -\bnabla_{\mu}
\bnabla_{\nu} +\frac{2 \Lambda_{eff}}{d-2} \bg_{\mu \nu}
\right)R_L+\alpha_3 \left(\bar{\Box} \G^L_{\mu \nu}-\frac{2
\Lambda_{eff}}{d-2}\bg_{\mu \nu} R_L \right)
\end{eqnarray}
where
\begin{equation}
\G^L_{\mu \nu} \equiv R^L_{\mu \nu}-\frac{1}{2}\bg_{\mu \nu}
R_L-\frac{2 \Lambda_{eff}}{d-2} h_{\mu \nu}\,,
\end{equation}
$R^L_{\mu \nu}$ is the linearized Ricci tensor, and $R_L$ is the
linearized Ricci scalar.  The coefficients are given by
\begin{eqnarray}
\alpha_1 &=& 1+\frac{4 d \Lambda_{eff}  b_1}{d-2}+\frac{4  (d-3)(d-4) \Lambda_{eff} b_2}{(d-2)(d-1)}+\frac{4 \Lambda_{eff} b_3}{d-1}-\frac{48 (2d-3)\Lambda_{eff}^2 c_1}{(d-2)^2(d-1)^2}+\frac{36\Lambda_{eff}^2 c_2}{(d-2)(d-1)^2} \nonumber \\
\alpha_2 &=& 2 b_1 +b_3 +\frac{24 \Lambda_{eff} c_1}{(d-2)(d-1)} \\
\alpha_3 &=& b_3+\frac{48 \Lambda_{eff} c_1}{(d-2)(d-1)}-\frac{6
\Lambda_{eff} c_2}{(d-2)(d-1)}\,. \nonumber
\end{eqnarray}
Remarkably, this result has precisely the same tensor form as the
four-derivative case; the only effect of the six-derivative terms is
to modify the coefficients $\alpha_i$.  Hence, we may simply borrow
the results of \cite{Deser:2002jk} to obtain the potential
\begin{eqnarray}
\label{potential}
16\,\pi\,G \mathcal{F}^{\nu \mu}&=&  \frac{\tilde \alpha_1}{2} \left[\bxi_\lambda \bnabla^\mu h^{\nu \lambda}-\bxi_\lambda \bnabla^\nu h^{\mu \lambda}+\bxi^\mu \bnabla^\nu h-\bxi^\nu \bnabla^\mu h +h^{\mu \lambda} \bnabla^\nu \bxi_\lambda -h^{\nu \lambda} \bnabla^\mu \bxi_\lambda +\bxi^\nu \bnabla_\lambda h^{\mu \lambda}-\bxi^\mu \bnabla_\lambda h^{\nu \lambda}+h \bnabla^\mu \bxi^\nu\right]\nonumber \\
&&+\alpha_2 \left[\bxi^\mu \bnabla^\nu R_L-\bxi^\nu \bnabla^\mu R_L-R_L \bnabla^\nu \bxi^\mu\right] +\alpha_3 \left[\bxi_\lambda \bnabla^\nu \G_L^{\mu
\lambda}-\bxi_\lambda \bnabla^\mu \G_L^{\nu \lambda}+\G_L^{\nu
\lambda}\bnabla^\mu\bxi_\lambda-\G_L^{\mu \lambda}\bnabla^\nu\bxi_\lambda \right]
\end{eqnarray}
with
\begin{equation}
\tilde \alpha_1 = 1+\frac{4 \Lambda_{eff} (d b_1+b_3)}{d-2}+\frac{4
(d-3)(d-4)\Lambda_{eff} b_2}{(d-2)(d-1)}-\frac{48  (2d-7)
\Lambda_{eff}^2 c_1}{(d-2)^2(d-1)^2}+\frac{12(3d-8) \Lambda_{eff}^2
c_2}{(d-2)^2(d-1)^2} \,.
\end{equation}
Note that at this stage, the expression for the energy is
exact in the couplings.

Consider a static, spherically symmetric, asymptotically AdS black
hole solution of the form
\begin{equation}
\label{sads} ds^2=-f_1(r)
dt^2+\frac{dr^2}{f_2(r)}+r^2d\Omega^2_{d-2,k},
\end{equation}
where $d\Omega^2_{d-2,k}$ is the line element of the transverse
space with curvature $k=0,1$, and as $r \to \infty$, we have
\begin{equation}
\label{sadsf2}
f_2(r)=\frac{r^2}{\ell_{eff}^2}+k+\frac{m}{r^{d-3}}+\ldots \,.
\end{equation}
Here $\ell_{eff}$ is the (effective) AdS radius, which is related to
the cosmological constant through
\begin{equation}
\ell_{eff}^2=-\frac{(d-2)(d-1)}{2 \Lambda_{eff}} \,.
\end{equation}
As noted in \cite{Deser:2002jk} for four-derivative gravity, it is
interesting that for the class of metrics \eqref{sads},
\eqref{sadsf2}, the $\alpha_2,\alpha_3$ terms in \eqref{potential}
fall off fast enough at infinity so that they give zero contribution
to the energy.   Hence, the only contribution to the energy is from
the first term, which is simply the Einstein-Hilbert result with a
coefficient corrected by the higher derivative terms. We now see
that the same is true for our six-derivative theory \eqref{action2}.
Using \eqref{esurf}, \eqref{potential} and \eqref{lambdapert}, the
final expression for the energy density (to order $\alpha'^2$) is
\begin{eqnarray}
\label{energy}
\mathcal{E} &=& \left(1+\frac{4 d   \Lambda \,b_1}{d-2}+\frac{4   \Lambda \,b_3}{d-2}+\frac{4 (d-4) (d-3)  \Lambda \,b_2}{(d-2) (d-1)}-\frac{48 (2 d-7)  \Lambda^2 \,c_1}{(d-2)^2 (d-1)^2}+\frac{12 (3 d-8)  \Lambda^2 \,c_2}{(d-2)^2 (d-1)^2}\right.\nonumber \\
&&-\frac{40 d \Lambda^2 \,b_1^2}{9 (d-2)}-\frac{8 \Lambda^2
\,b_3^2}{9 (d-2)}
-\frac{4 (d-4) (d-3) \Lambda^2 \,b_2^2}{3 (d-2) (d-1)}-\frac{4 \left(13 d^2-73 d+120\right) \Lambda^2 \,b_1  b_2 }{9 (d-2)(d-1)}\nonumber \\
&&-\frac{8 (d+5) \Lambda^2 \,b_1  b_3}{9 (d-2)}-\frac{4 \left(2
d^2-11 d+21\right) \Lambda^2 \,b_3  b_2}{9 (d-2) (d-1)} +\ldots
\Big) \mathcal{E}_0
\end{eqnarray}
where $\mathcal{E}_0$ is the result for Einstein-Hilbert gravity
\begin{equation}
\label{ehenergy} \mathcal{E}_0=\frac{(d-2)\,m}{32\,\pi\,G} \,.
\end{equation}
Note that the expression for the energy (\ref{energy}) is perturbative, since it was obtained using the perturbative solution for $\Lambda_{eff}$,
(\ref{lambdapert}). This is the relevant
form for our analysis since we treat the higher derivative terms
as corrections to the Einstein-Hilbert action.
The leading $\alpha'$ corrections in \eqref{energy} match exactly
the result of \cite{Cremonini:2009ih} for four-derivative gravity,
which was obtained through boundary counterterm methods.  While the
counterterm results were only strictly valid for $d<7$, the result
\eqref{energy} confirms the expectation that the expression for the
energy in \cite{Cremonini:2009ih} holds in all $d$.


\subsection{Charged AdS$_{5}$ Black Branes}
\label{mq5d}

To address the WGC, we want to consider a charged black brane in the
theory \eqref{action2}.  The simplest way to add charge is to choose
the matter sector to contain just a Maxwell field with the minimal
term
\begin{equation}
L_m = -\frac{1}{4} F^2  \quad \Rightarrow \quad \tau_{\mu \nu}=
\frac{1}{32\,\pi\,G}\left(F_{\mu \lambda}
F_{\nu}{}^\lambda-\frac{1}{4}F^2 g_{\mu \nu} \right)\,,
\end{equation}
where $F_{\mu \nu}=\nabla_\mu A_\nu -\nabla_\nu A_\mu$ and $F^2 =
F_{\mu \nu} F^{\mu \nu}$. Hence, even though there are higher
curvature terms, the matter equation of motion is still
\begin{equation}
\nabla^\mu F_{\mu \nu}=0
\end{equation}
and the charge is given by the usual expression
\begin{eqnarray}
\label{charge} \mathcal{Q} = \int_{\partial\Sigma}
d^{d-2}x\sqrt{g_{\partial\Sigma}} \,n_\mu r_\nu F^{\mu \nu} \,.
\end{eqnarray}

Let us now restrict to planar ($k=0$) black branes in $d=5$ and work
in units where the uncorrected AdS radius is set to $\ell =1$.  We
consider a general ansatz of the form
\begin{eqnarray}
\label{ansatz1}
ds^2 &=&-\omega(r) dt^2+\frac{\sigma^2(r)}{\omega(r)}dr^2+r^2\left(dx^2 +dy^2 +dz^2\right) \\
\label{ansatz2} A_\mu &=& \gamma(r) \delta_{\mu}^t \,.
\end{eqnarray}
In the absence of higher derivative corrections, the solution to the
equations of motion is given by
\begin{eqnarray}
\omega_{(0)}(r) &=& r^2\(1-\frac{r_0^2}{r^2}\)\(1+\frac{r_0^2}{r^2}-\frac{q^2}{r_0^2\,r^4}\) \label{leading} \\
\sigma_{(0)}(r) &=& 1  \nonumber \\
\gamma_{(0)}(r) &=& \frac{\sqrt{3} \,q }{r^2} \,. \nonumber
\end{eqnarray}
We assume $q^2 \leq 2 r_0^6$, so that $r = r_0$ corresponds to the
outer horizon. The Hawking temperature of a black brane of the type
\be ds^2=g_{tt}(r)dt^2+g_{rr}(r)dr^2+g_{xx}(dx^2+...) \ee is given
by the general formula \be \label{temperature}
T=-\left.\frac{\partial_{r}g_{tt}}{4\,\pi\,\sqrt{-g_{tt}\,g_{rr}}}\right|_{r\rightarrow
r_0}=\frac{\omega'(r_0)}{4\,\pi\,\sigma(r_0)}, \ee which for the
above solution becomes
\begin{equation}
T_{(0)}= \frac{r_0}{\pi}\left(1- \frac{q^2}{2r_0^6} \right) \,.
\end{equation}
The extremal solution is defined by the condition $T =0$, which
corresponds to $q^2 = 2 r_0^6$. This is also the point where the
outer and inner horizons coincide.  The mass is given by
\eqref{ehenergy} with
\begin{equation}
m = \frac{q^2+r_0^6}{r_0^2}
\end{equation}
and using \eqref{charge}, the charge density is
\begin{equation}
\label{bhcharge} Q = 2 \sqrt{3} \,q \,.
\end{equation}

Using the ansatz \eqref{ansatz1}, \eqref{ansatz2} and the leading
order solution, we can solve the equations of motion perturbatively
in $\alpha'$.  The results are given in appendix
\ref{CorrectedSolution}, using a scheme where the horizon radius is
fixed at $r=r_0$, i.e.~it is not corrected by the higher derivative
contributions.  This can be achieved by choosing the integration
constants appropriately when solving the gravitational field
equations at each order in $\alpha'$.  This does, however, produce
corrections to the $O(r^{-2})$ ``mass term'' in the metric, so the
$m$ in \eqref{energy} should properly be viewed as a function of the
parameters $q, r_0, b_i, c_i$. The cosmological constant gets
corrected as given in \eqref{lambdapert}.  We also choose
integration constants when solving Maxwell's equations so that the
charge is not corrected and remains as in \eqref{bhcharge}. The
Hawking temperature of the corrected solution is also given in
appendix \ref{CorrectedSolution}.

 To set the speed of light to be one in the dual CFT, one
should actually rescale the time coordinate by a red-shift factor
$t\to t/\ell_{eff}$.  Equivalently, as noted in
\cite{Cremonini:2009ih}, we obtain the physical energy density of
the field theory by simply rescaling $M \equiv \ell_{eff}
\mathcal{E}$. Similarly, the temperature of the CFT is given by
$T_{CFT}\equiv \ell_{eff} T$.  Now, we want to compare thermodynamic
quantities (like $M/Q$) of the uncorrected solution to those of the
corrected solution\footnote{Note that we take the ratio of two
densities, so that the volume factors cancel.}. It is important to
remember that this is meaningful only if the temperature does not
change when the higher derivative terms are turned on.  Thus, we
would actually like to write $M/Q$ in terms of $T_{CFT}$ instead of
$r_0$.  To do so, we introduce a new parameter $\bar r_0$ and fix
\begin{equation}
\label{rbar} T_{CFT} = \frac{\bar r_0}{\pi}\left(1-
\frac{q^2}{2\bar r_0^{\,6}} \right)\,.
\end{equation}
This relation may be solved (perturbatively) to give $r_0 = r_0(\bar
r_0, q)$ so that we may eliminate $r_0$ in favor of $\bar r_0$ in
all expressions. The explicit expression for $\bar{r}_0$ is given in
appendix \ref{CorrectedSolution}. Then $\bar r_0$ is implicitly a
function of $(T_{CFT},q)$ through \eqref{rbar} and we have ensured
that the corrected and uncorrected solutions have the same
temperature.   The extremal case is given precisely by $q^2 = 2 \bar
r_0^6$.

Hence, the result for the mass to charge ratio is
\begin{eqnarray}
\label{mq} \frac{M}{Q} &=& \left(\frac{M}{Q}\right)_0 \left(1+\frac{
\tQ^6-186 \tQ^4-60
   \tQ^2-200}{2 \left(\tQ^2+1\right) \left(5 \tQ^2+2\right)}b_1-\frac{3  \left(\tQ^4+5 \tQ^2-2\right)}{\left(\tQ^2+1\right) \left(5\tQ^2+2\right)}b_2\right.
\nonumber \\
&&+\frac{ 11 \tQ^6-102 \tQ^4-12 \tQ^2-40}{2\left(\tQ^2+1\right)
\left(5 \tQ^2+2\right)} b_3
-\frac{16602 \tQ^8-59267 \tQ^6+23548 \tQ^4-7798 \tQ^2+5180}{35 \left(\tQ^2+1\right) \left(5 \tQ^2+2\right)}c_1\nonumber \\
&&-\frac{3  \left(846 \tQ^8-4521 \tQ^6+6664 \tQ^4-574
\tQ^2-4340\right)}{140 \left(\tQ^2+1\right)
\left(5\tQ^2+2\right)}c_2-\frac{27  \left(\tQ^2+2\right) \left(15
\tQ^6+39 \tQ^4-16 \tQ^2-4\right)}{2 \left(\tQ^2+1\right)
\left(5\tQ^2+2\right)^3}b_2^2
\nonumber \\
&&+\frac{ 1860 \tQ^{12}-7445 \tQ^{10}-98484 \tQ^8-154414
\tQ^6-272756 \tQ^4+13800\tQ^2+2000}{3 \left(\tQ^2+1\right) \left(5
\tQ^2+2\right)^3}b_1^2
\nonumber \\
&&+\frac{ 4020 \tQ^{12}+5335\tQ^{10}-143088 \tQ^8-155962
\tQ^6-144956 \tQ^4+3864 \tQ^2+560}{21 \left(\tQ^2+1\right)
\left(5\tQ^2+2\right)^3}b_3^2
\nonumber \\
&&+\frac{ 1200 \tQ^{12}-7165\tQ^{10}-24978 \tQ^8-211352 \tQ^6-9712
\tQ^4-47568 \tQ^2-800}{10 \left(\tQ^2+1\right)
\left(5\tQ^2+2\right)^3}b_1 b_2
\nonumber \\
&&+\frac{4 \left(2760 \tQ^{12}+2605 \tQ^{10}-156759 \tQ^8-201175
\tQ^6-260330 \tQ^4+9660\tQ^2+1400\right)}{21 \left(\tQ^2+1\right)
\left(5\tQ^2+2\right)^3}b_1 b_3
\nonumber \\
&&\left.+\frac{1200 \tQ^{12}-6085 \tQ^{10}-8202 \tQ^8-84416
\tQ^6-11872 \tQ^4-18480\tQ^2-224}{14 \left(\tQ^2+1\right) \left(5
\tQ^2+2\right)^3}b_2 b_3 +\ldots\right) \,,
\end{eqnarray}
where we have set $\tQ \equiv q/\bar r_0^{\,3}$
 and
\begin{equation}
\left(\frac{M}{Q}\right)_0 =\frac{\sqrt{3} \left(1+\tQ^2\right) \bar
r_0}{64\,\pi\,G \tQ} \,.
\end{equation}

At extremality, the mass to charge ratio becomes
\begin{eqnarray}
\label{mqex}
\frac{M}{Q} &=& \left(\frac{M}{Q}\right)_0 \left(1-\frac{44}{3} b_1- b_2 -\frac{16}{3}b_3 +\frac{10394 }{105 } c_1+\frac{61 }{70 }c_2 -\frac{770}{3} b_1^2\right. \nonumber \\
&&\left.\qquad \qquad \quad  -\frac{5}{2} b_2^2-\frac{710}{21} b_3^2-\frac{3764}{21}
b_1 b_3-\frac{688}{15} b_1 b_2 -\frac{292 }{21 }  b_3 b_2+ \ldots
\right)
\end{eqnarray}
where
\begin{equation}
\left(\frac{M}{Q}\right)_0 = \frac{3 \sqrt{3} \, \bar r_0}{ \sqrt{2}
\, 64\,\pi\,G} \,.
\end{equation}
The corresponding result for black branes with spherical horizons is
given in appendix \ref{sphere}.

\section{The Shear Viscosity to Entropy Density Ratio}
\label{viscosity_section} In this section, we calculate perturbative
corrections to the shear viscosity to entropy density ratio of a CFT
plasma dual to the static charged black brane solution
(\ref{ansatz1},\ref{leading}). The corrections we consider are given
in (\ref{action}). We begin by giving a brief summary of the
holographic method for calculating the shear viscosity of the dual
CFT. We then compute the entropy density using Wald's formula for
the Noether charge entropy and use these results to obtain the
ratio.

\subsection{The Shear Viscosity}
To compute the viscosity, we use the prescription given
in~\cite{MyersChemical,Policastro,Son1,Dutta1}. We present
here only the main steps, mostly
following~\cite{MyersChemical}, where the four-derivative correction to
the viscosity with a chemical potential was computed.

The viscosity of the boundary field theory is given by Kubo's
formula:
\be
 \eta=\lim_{\omega\rightarrow 0}\frac{1}{2\omega}\int dt
d^3x\,e^{i\omega t}\langle\left[T_{xy}(t,x),T_{xy}(0,0)\right]
\rangle \,.
\ee
The two-point function in the formula above can be expressed as a
retarded Green's function of $T_{xy}$:
\be
G^{R}_{xy,xy}(\omega,\textbf{k})=-i\int\,d^4x\,e^{i(\omega
t-\textbf{k}\cdot \textbf{x})}\theta(t)\langle\left[T_{xy}(t,x),T_{xy}(0,0)\right]
\rangle
\ee
so that
\be \eta=-\lim_{\omega \rightarrow 0}
\frac{1}{\omega}\,\text{Im}\,G^{R}_{xy,xy}(\omega,0) \,.
\ee

The boundary operator $T_{xy}$ is coupled to a metric perturbation
(graviton) $h^{xy}$ in the bulk. Hence, we can perform the Green's
function computation on the metric perturbation
$g_{\mu\nu}\rightarrow g_{\mu\nu}+h_{xy}(t,r)$. It turns out that
the equation of motion of $h_{x}^{y}$ is that of a minimally coupled
scalar in the case of the Einstein-Hilbert action. Higher derivative
corrections enter as effective modifications of the coefficients of
this scalar equation. We define \be \phi(r,t,\textbf{x})\equiv
h_{x}^{y}(r,t,\textbf{x})=\int
\frac{d^{4}k}{(2\pi)^4}\phi_k(r)e^{-i\omega t+i\textbf{k}\cdot
\textbf{x}} \ee and expand the action to second order in
$\phi(r,t)$. This effective action is denoted by $I^{(2)}_{\phi}$.
Next by computing the radial canonical momentum \be
\Pi(r)\equiv\frac{\delta I^{(2)}_{\phi}}{\delta \phi_k'} \ee we find
the retarded Green's function \be G^{R}_{xy,xy}=-\lim_{r\rightarrow
\infty}\frac{\Pi(r)}{i \phi_k(r)}. \ee

For theories without derivatives of the curvature, the effective
action is of the general form \bea \label{effective}
I^{(2)}_{\phi}&=&\frac{1}{16\pi\,G}\int \frac{d^{4}k}{(2\pi)^4}\,
dr\(A(r)\phi_k''\phi_{-k}+B(r)\phi'_k\phi'_{-k}+C(r)\phi'_k\phi_{-k}\right.\\
&+&\left.
D(r)\phi_k\phi_{-k}+E(r)\phi''_k\phi''_{-k}+F(r)\phi''_k\phi'_{-k}\)+\textit{boundary
terms}. \nonumber
\eea
Note that while one technically requires the appropriate boundary terms so that the action is well-defined, it turns out that their explicit form is not required to obtain the viscosity.  The canonical conjugate momentum to $\phi$ is then
\be
\Pi(r)=\frac{1}{8\pi\,G}\left[\(B-A-\frac{1}{2}F'\)\phi'-\(E\phi''\)'\right].
\ee

Since in the limit $\omega \rightarrow 0$ the equation of motion is
\be
\partial_r\Pi=0,
\ee we can compute the Green's function at any point. In particular,
we demand infalling boundary conditions on $\phi(r)$ at the horizon
(to avoid a singularity there), namely \be
\left.\phi(r,t)\right|_{r=r0}=\phi(v) \,, \ee where $v$ is the
Eddington-Finkelstein coordinate defined as
$dv=dt+\sqrt{-\frac{g_{rr}}{g_{tt}}}dr$. Then we find
that~\cite{liu+iqbal} \be
\partial_r\phi(r,t)=\sqrt{-\frac{g_{rr}}{g_{tt}}}\partial_t\phi(r,t) \Rightarrow \partial_r\phi_k =  -i\omega\sqrt{-\frac{g_{rr}}{g_{tt}}}\phi_k
\ee gives a convenient formula for the
viscosity~\cite{MyersChemical}
 \bea
 \label{visc}\eta&=&\left.\lim_{\omega\rightarrow 0}\frac{\Pi(r)}{i\omega\phi(r)}\right|_{r=r_0}=\frac{1}{8\pi
 G}\left[\kappa_2(r_0)+\kappa_4(r_0)\right],
 \eea
 with
 \bea
 \kappa_2(r)&=&\sqrt{-\frac{g_{rr}}{g_{tt}}}\(A(r)-B(r)+\frac{F'(r)}{2}\),
 \qquad \qquad \kappa_4(r)=\(E(r)\(\sqrt{-\frac{g_{rr}}{g_{tt}}}\)'\)'.
 \eea

Now we are ready to apply this prescription to the action
(\ref{action}). We calculate the effective action (\ref{effective})
for the perturbation $\phi$  using the metric ansatz (\ref{ansatz1})
as the background metric. The effective coefficients in the action
$A(r),B(r),E(r),F(r)$ are given in appendix \ref{expVisco}.
Substitution into the formula (\ref{visc}) gives \bea
\label{viscexp1} \non
\eta&=&\frac{r^3_0}{16\,\pi\,G}\(1+\frac{2\,b_1}{\sigma(r_0)^3}\left[\(\sigma'(r_0)
-\frac{6\,\sigma(r_0)}{r_0}\)\omega'(r_0)-\sigma(r_0)\,\omega''(r_0)\right]
+\frac{b_3}{\sigma(r_0)^3}\left[\(\sigma'(r_0)-\frac{3\,\sigma(r_0)}{r_0}\)\omega'(r_0)
\right. \right.
 \\ && \nonumber \left.\left.
-\sigma(r_0)\,\omega''(r_0)\right]\right.-\frac{2\,b_2\omega'(r_0)}{r_0\,\sigma(r_0)^2}
-\frac{6\,c_1\,\omega'(r_0)}{r_0^2\,\sigma(r_0)^5}\left[\(\sigma(r_0)+4\,r_0\,\sigma'(r_0)\)\omega'(r_0)-2\,r_0\,\sigma(r_0)\,\omega''(r_0)\right]
\\
&&-\frac{3\,c_2}{4\,r_0^2\,\sigma(r_0)^6}\left[\omega'(r_0)^2\(4\,\sigma(r_0)^2\right.\right.\left.
+3\,r_0\,\sigma(r_0)\,\sigma'(r_0)+4\,r_0^2\sigma'^2(r_0)-r_0^2\,\sigma(r_0)\,\sigma''(r_0)\)
\non
\\ &&\left.\left. +r_0\omega'(r_0)\(r_0\,\sigma^2(r_0)\,\omega^{(3)}(r_0)
-\(5\,r_0\,\sigma(r_0)\,\sigma'(r_0)+3\,\sigma^2(r_0)\)\,\omega''(r_0)\)+r_0^2\,\sigma(r_0)^2\,\omega''^2(r_0)\right]\).
 \eea

In principle, it is possible to compute the correction to the
viscosity based only on the near horizon solution as demonstrated
in~\cite{Dutta2}. However, since we already required the full
solution in section \ref{mass_section}, we may just use the metric
corrections given explicitly in appendix~\ref{CorrectedSolution}.
The leading order expressions are given in eq.~(\ref{leading}).
Substitution in (\ref{viscexp1}) then yields the shear viscosity in
terms of the charge and the horizon radius: \bea \label{visc_final}
\eta&=&\frac{r^3_0}{16\,\pi\,G}\left[1-4\,b_1\(10+\tilde{q}^2\)-4\,b_2\(2-\tilde{q}^2\)-8\,b_3\(1+\tilde{q}^2\)
-\frac{8}{3}\,b_1^2\(97\,\tilde{q}^4-172\,\tilde{q}^2+100\)\right.\\
\non
&&-\frac{8}{3}\,b_3^2\(19\,\tilde{q}^4+32\,\tilde{q}^2+4\)-\frac{8}{3}\,b_{2}\,b_3\(7\,\tilde{q}^4-64\,\tilde{q}^2+28\)
-\frac{16}{3}\,b_1\,b_3\(41\tilde{q}^4-2\tilde{q}^2+20\)\\
\non && \left.
-\frac{8}{3}\,b_1\,b_2\(17\,\tilde{q}^4-140\,\tilde{q}^2+68\)-72\,c_1\(5\tilde{q}^4-12\tilde{q}^2+4\)-6\,c_2\(61\,\tilde{q}^4-100\,\tilde{q}^2+28\)\right],
\eea where $\tilde{q}\equiv q/r_0^3$ is the dimensionless charge
parameter.


\subsection{The Entropy Density}

For the computation of the entropy density, we use Wald's formula for Noether charge density~\cite{wald1,wald2}.
Wald's formula is consistent with the
first law of thermodynamics and therefore also with the Euclidean
approach. The Noether charge entropy is given in the form of an integral over
fields on a spatial section of the horizon $\mathcal{H}$. For theories
without derivatives of the Riemann tensor, Wald's formula takes the
following form~\cite{myers}:
\be
S_{BH}=-2\,\pi\,\int_{\mathcal{H}}\frac{\partial\,L}{\partial R_{\mu \nu
\rho \sigma}}\,\epsilon_{\mu\nu}\,\epsilon_{\rho\sigma}\,
\sqrt{g_\mathcal{H}}\,d\Omega_{d-2}~,
\ee
where the action of the $d$-dimensional
theory is
\be
I=\int\!d^{d}x\sqrt{-g}\,L~,
\ee
and $\epsilon_{\mu\nu}$ is the binormal to the spatial section of the
horizon $\mathcal{H}$, i.e. the volume element orthogonal to it. The binormal
is defined by $\epsilon_{\mu\nu}=\nabla_{\mu}\chi_{\nu}$,
where $\chi_{\nu}$ is a Killing field normalized so that
$\epsilon_{\mu\nu}\epsilon^{\mu\nu}=-2$. The volume element induced on $\mathcal{H}$ is denoted $\sqrt{g_\mathcal{H}}\,d\Omega_{d-2}$. For black branes of the
form~(\ref{ansatz1}), the integration in the spatial directions $x,y,z$
gives an infinite factor. Therefore we consider only the entropy
density $s$ in those directions.
Substitution of the corrected solution which appears in appendix \ref{CorrectedSolution}
into the expression given by Wald's formula gives us the entropy density as
\bea
\label{entropyeq}
s&=&\frac{r_0^3}{4\,G}\left[1-4\,b_1\(10+\tilde{q}^2\)-8\,b_3\(1+\tilde{q}^2\)-\frac{8}{3}\,b_1^2\,\(97\,\tilde{q}^4-172\,\tilde{q}^2+100\)-\frac{8}{3}\,b_3^2\,\(19\,\tilde{q}^4+32\,\tilde{q}^2+4\)\right.\\
\non && \left.
-16\,b_2\(2\,b_1+b_3\)\(\tilde{q}^2-2\)^2-\frac{16}{3}\,b_1\,b_3\,\(41\,\tilde{q}^4-2\,\tilde{q}^2+20\)+12\,c_1\(7\,\tilde{q}^2-2\)^2+9\,c_2\(\tilde{q}^2-2\)
\right] \,. \eea

\subsection{The Ratio}
Combining \eqref{entropyeq} with the result for the shear viscosity
(\ref{visc_final}) and rewriting the expressions with $\bar{r}_0$
using eq. (\ref{rbar2}), we find that the shear viscosity to entropy
density ratio is given by \bea \label{theratio}
\frac{\eta}{s}&=&\frac{1}{4\pi}\left[
1-4\,b_2\(2-\bq^2\)-\frac{24\,b_2^2\,\bq^2\,\(2-\bq^2\)}{2+5\,\bq^2}+\frac{8\,b_1\,b_2\,\(2-\bq^2\)\,\(49\bq^4-250\,\bq^2-140\)}{3\,(2+5\,\bq^2)} \right. \\
\non && \left.
-\frac{8\,b_2\,b_3\,\(2-\bq^2\)\,\(\bq^4+50\,\bq^2+28\)}{3\,(2+5\,\bq^2)}
-12\,c_1\,\(79\bq^4-100\,\bq^2+28\)-3\,c_2\,\(125\,\bq^4-212\,\bq^2+68\)
\right]\,.\eea In the extremal limit $\bar{q}^2\rightarrow 2$, the
ratio becomes
 \be
 \label{extrat}
\frac{\eta}{s}= \frac{1}{4\pi}\left[1-432\(4\,c_1+c_2\)\right]\,.
\ee
Note that all contributions from the four-derivative terms vanish at the
extremal limit, while the six-derivative corrections survive.


\section{Constraints from the WGC and the KSS Bound}
\label{analysis} In this section we analyze the results for $M/Q$
and $\eta/s$ to determine the conditions under which the WGC and the
KSS bound are compatible.  Given the number of parameters in the
action \eqref{action}, it is convenient to discuss various cases in
turn.

\subsection{Gauss-Bonnet}
\label{gbbounds}

\begin{figure}[htb]
\begin{picture}(0,0)
\put(-150,-135){$\tQ$} \put(0,-275){$b_2$}
\end{picture}
\begin{center}
\includegraphics[width=3.5in]{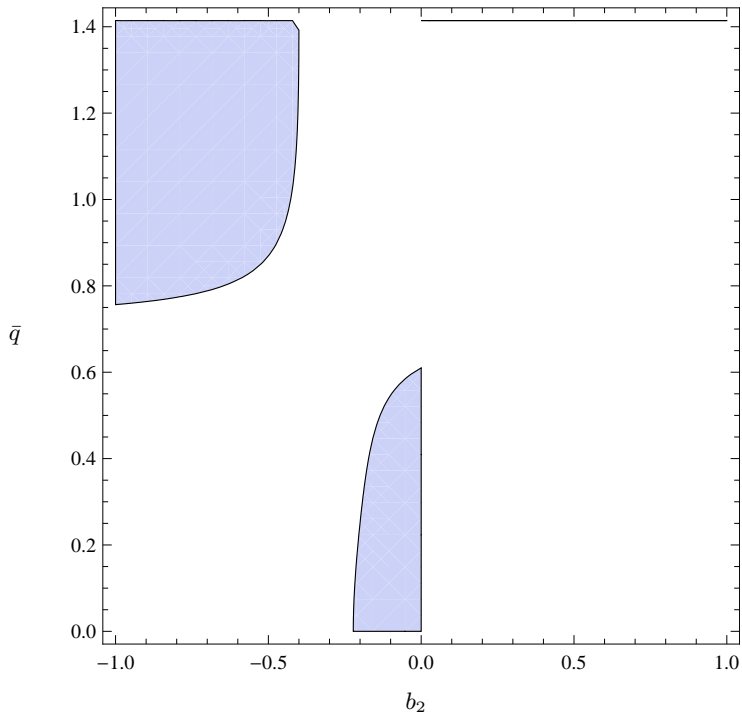}
\end{center}
\caption{Here we consider a theory with $b_1 = b_3 = c_1 = c_2 = 0$,
i.e. we keep only the Gauss-Bonnet term in the gravitational action.
The shaded regions represent points in the $(b_2, \tQ)$ plane where
both the WGC and the KSS bound are satisfied
simultaneously. Note that the disconnected line segment in the upper right
corner for $b_2>0$ corresponds to satisfaction of both
constraints exactly at extremality.  Since this case is disconnected from
the non-extremal region and thus non-generic, it is not expected to occur physically.} \label{gbplot}
\end{figure}

We first consider theories in which the six-derivative terms are
turned off, as would be required by supersymmetry for example.  As
we have seen above, $O(\alpha'^2)$ corrections to $M/Q$ and $\eta/s$
still arise due to terms quadratic in the four-derivative couplings.
We begin by choosing a theory with $b_1 = b_3 = c_1 = c_2 = 0$.
This combination of the curvature-squared Gauss-Bonnet term and the
absence of curvature-cubed terms arises, for example, in the low
energy effective action of the heterotic string (see e.g.~\cite{Tseytlin1}).

At extremality, it follows from \eqref{mqex} and (\ref{extrat}) that
\begin{equation}
\frac{M}{Q} = \left(\frac{M}{Q}\right)_0 \left(1-b_2 -\frac{5}{2}
b_2^{\,2}\right) \,, \quad \frac{\eta}{s} = \frac{1}{4 \pi} \,.
\end{equation}
The requirement that the $\alpha'$ corrections reduce $M/Q$ implies
$b_2 <-2/5$ or $b_2 >0$.  In fact, both bounds are satisfied in this
range, as the KSS bound clearly holds for any $b_2$.

\begin{figure}[htb]
\begin{picture}(0,0)
\put(-150,-135){$\tQ$} \put(0,-275){$b_2$}
\end{picture}
\begin{center}
\includegraphics[width=3.5in]{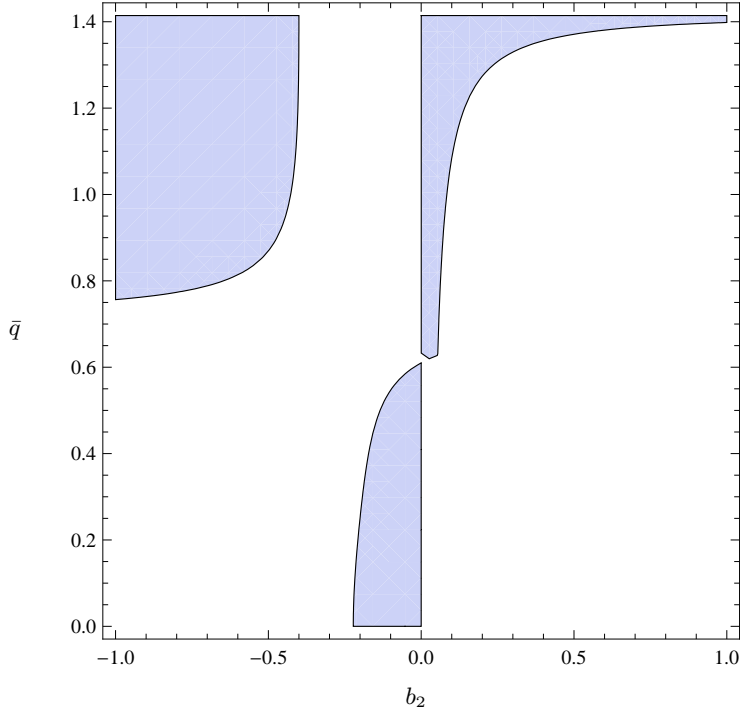}
\end{center}
\caption{Here we again consider only the Gauss-Bonnet correction, but instead relax the lower bound on $\eta/s$ to be $4/25 \pi$.
The shaded regions represent points in the $(b_2, \tQ)$ plane where
both the WGC and the (improved) viscosity bound are satisfied
simultaneously. Compared to Figure \ref{gbplot}, there is now a nontrivial region near extremality for $0 < b_2 \ll 1$ where both constraints hold.} \label{gbplot2}
\end{figure}

Since we expect that the WGC should hold also
for a neighborhood of the extremal limit, we now analyze the constraints for non-extremal black holes. The result \eqref{mq} implies that
for non-extremal black holes near extremality, the WGC is satisfied for $b_2 >0$ or
\begin{equation}
\label{gbwgc} b_2 <-\frac{2 \left(5 \tQ^2+2\right)^2 \left(\tQ^4+5
\tQ^2-2\right)}{9 \left(15 \tQ^8+69 \tQ^6+62 \tQ^4-36
\tQ^2-8\right)} \leq -\frac{2}{5} \,.
\end{equation}
The result (\ref{theratio}) shows that the KSS bound holds (for all $\tQ$)
when $b_2 <0$, so near extremality both bounds are satisfied when
\eqref{gbwgc} holds.  The regions in the $(b_2, \tQ)$ plane where
the two bounds are compatible are plotted in Figure \ref{gbplot}.
Note that for the heterotic string theory effective action in
particular, the Gauss-Bonnet coupling is positive~\cite{Tseytlin1}, so in this case both bounds cannot be satisfied away
from the extremal limit.  However, if we instead consider the viscosity bound of $4/25 \pi$ \cite{causality}, the region of ``good'' theories is enlarged and does include cases with $b_2 >0$ (see Figure \ref{gbplot2}).

\subsection{Weyl-Tensor-Squared}
\label{w2bounds}

\begin{figure}[htb]
\begin{picture}(0,0)
\put(-150,-140){$\tQ$} \put(0,-275){$b_2$}
\end{picture}
\begin{center}
\includegraphics[width=3.5in]{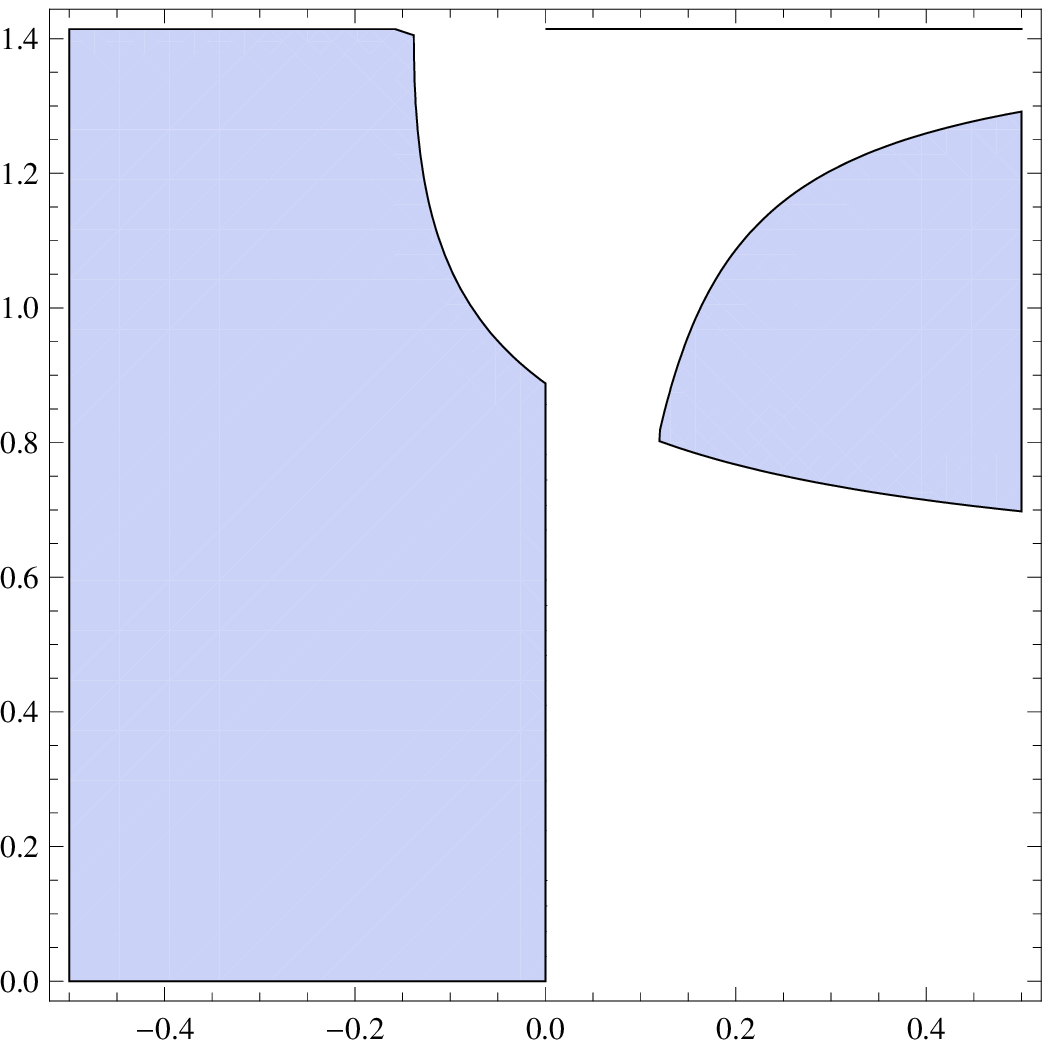}
\end{center}
\caption{Here we consider a theory with $b_1 =-5 b_2/6, b_3 = 8
b_2/3, c_1 = c_2 = 0$, which corresponds to keeping only the
Weyl-tensor-squared term in the gravitational action.  The shaded
regions represent points in the $(b_2, \tQ)$ plane where both the
WGC and the KSS bound are satisfied
simultaneously.  See Figure \ref{gbplot} for comments about the disconnected line segment in the upper right corner.}
\label{w2plot}
\end{figure}

As a second example of a four-derivative theory, we consider the
Weyl-tensor-squared theory given by setting $b_1 =-5 b_2/6, b_3 = 8
b_2/3, c_1 = c_2 = 0$.  A term of this form is present in higher
derivative corrections to $\mathcal{N} =2, d= 5$ gauged
supergravity~\cite{Cremonini}. Such supergravity theories  may arise
from compactifying type IIB string theory on $AdS_5 \times X^5$,
where $X^5$ is a Sasaki-Einstein manifold.  According to the AdS/CFT
correspondence, this theory is dual to $\mathcal{N} =1, d= 4$ Super
Yang Mills.

At extremality, it follows from \eqref{mqex} and (\ref{extrat}) that
\begin{equation}
\frac{M}{Q} = \left(\frac{M}{Q}\right)_0 \left(1-3 b_2
-\frac{152}{7} b_2^{\,2}\right) \,, \quad \frac{\eta}{s} =
\frac{1}{4 \pi} \,.
\end{equation}
The requirement that the $\alpha'$ corrections reduce $M/Q$ implies
$b_2 <-21/52\approx-0.14$ or $b_2 >0$.  In fact, both bounds are
satisfied in this range, as the KSS bound clearly holds for any
$b_2$.

For general $\tQ$ in the region near extremality, the WGC is satisfied for $b_2 >0$ or
\begin{equation}
\label{w2wgc} b_2 <-\frac{7 \left(5 \tQ^2+2\right)^2 \left(19
\tQ^6-82 \tQ^4-8 \tQ^2+48\right)}{4 \left(1755 \tQ^{12}-3690
\tQ^{10}-10688 \tQ^8-6536 \tQ^6-10080
   \tQ^4+3136 \tQ^2+448\right)} <0\,.
\end{equation}
The KSS bound is satisfied when $b_2 <0$ or
\begin{equation}
\label{w2visc} b_2> -\frac{5 \tQ^2+2}{29 \tQ^4-44 \tQ^2-28}>0 \,.
\end{equation}
Thus, near extremality both bounds are satisfied when \eqref{w2wgc}
or \eqref{w2visc} holds.  The regions in the $(b_2, \tQ)$ plane
where the two bounds are compatible are plotted in Figure
\ref{w2plot}.  For $b_2$ sufficiently small that the $O(b_2^2)$
corrections can be neglected, the two bounds are incompatible for
non-extremal cases, as pointed out in \cite{Cremonini:2009ih}.
However, we see that for $b_2 \lesssim -0.14$ the contribution from
the $O(b_2^2)$ correction allows both bounds to be satisfied
simultaneously.

For Weyl-tensor-squared theories, the behavior when the viscosity
bound is relaxed to $4/25 \pi$ is qualitatively similar to that
discussed in the previous subsection.


\subsection{Six-Derivatives}
\label{6bounds}

\begin{figure}[htb]
\begin{picture}(0,0)
\put(-124,-40){$\tQ = \sqrt{2}$} \put(24,-40){$\tQ = 1.3$}
\put(174,-40){$\tQ = 1.2$} \put(-124,-187){$\tQ = 1$}
\put(24,-187){$\tQ = 0.8$} \put(174,-187){$\tQ = 0.4$}
\put(-225,-75){$c_2$} \put(-140,-150){$c_1$}
\end{picture}
\begin{center}
\includegraphics[width=6in]{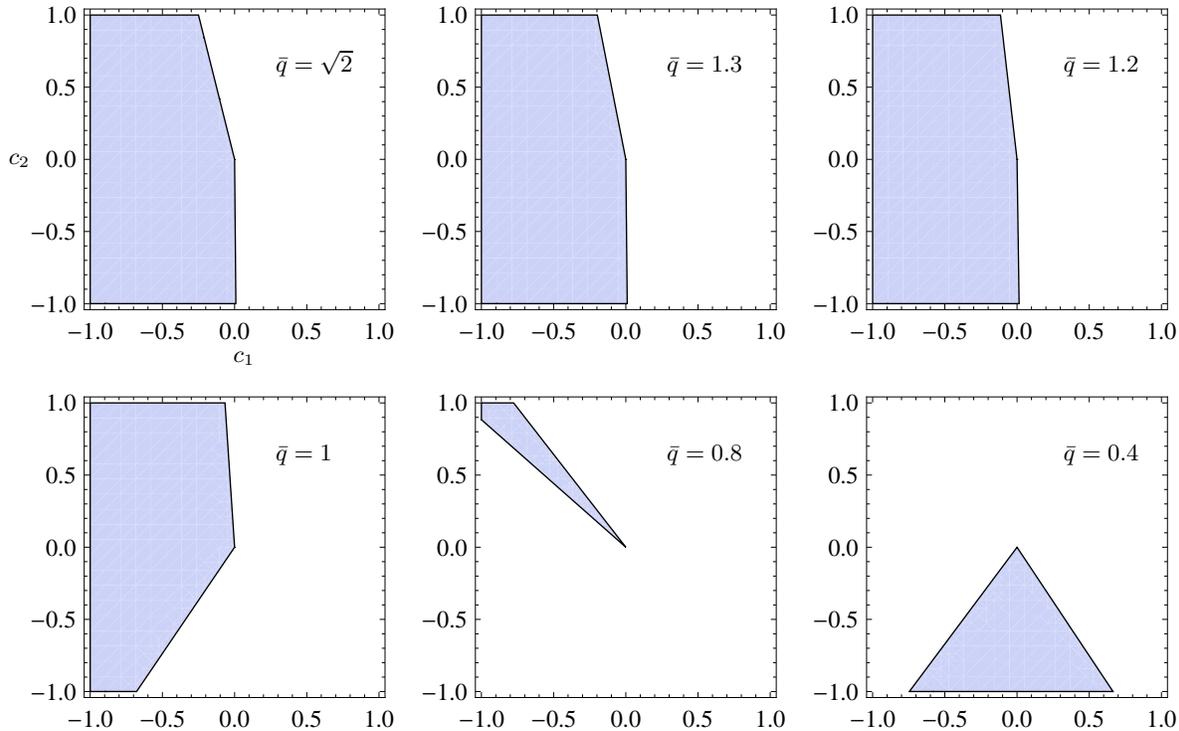}
\end{center}
\caption{Here we consider a theory with $b_1 = b_2= b_3 = 0$, i.e.
we keep only the six-derivative terms in the gravitational action.
The shaded regions represent points in the $(c_1, c_2)$ plane (for
various values of $\tQ$) where both the WGC and
the KSS bound are satisfied simultaneously.} \label{c1c2plot}
\end{figure}

We now consider theories in which the first corrections to the
effective action involve six-derivatives, i.e., $b_1 = b_2= b_3 =
0$. Such theories may serve as toy models for non-supersymmetric
string theory compactifications, which may in turn be dual to CFTs
with broken supersymmetry~\cite{Myers2}. Since we do not know in
general whether $c_1$ and $c_2$ are related, we will consider all
possibilities.

First suppose $c_2 = 0$.  Then the WGC implies that $c_1 < 0$ for $
0.73 \lesssim |\tQ| \leq \sqrt{2}$ and  $c_1 > 0$ for $ 0 < |\tQ|
\lesssim 0.73$.  Imposing the KSS bound requires $c_1 < 0$ for $
0.92 \lesssim |\tQ| \leq \sqrt{2}$ or $ 0 < |\tQ| \lesssim 0.65$,
and $c_1 >0$ for $ 0.65 \lesssim |\tQ| \lesssim 0.92$.  Thus, both
bounds can be satisfied when $ 0.92 \lesssim |\tQ| \leq \sqrt{2}$
with $c_1<0$ (which includes the extremal case) and $ 0.65 \lesssim
|\tQ| \lesssim 0.73$ with $c_1>0$.

Now suppose $c_1 = 0$.  Then the WGC is satisfied for all $ 0 <
|\tQ| \leq \sqrt{2}$ when $c_2 <0$. This condition is compatible
with the KSS bound when $ 1.12 \lesssim |\tQ| \leq \sqrt{2}$ (which
includes the extremal case) and $ 0 < |\tQ| \lesssim 0.66$.

Finally, we consider both $c_1, c_2 \neq 0$ and focus on the region
near extremality.  We have
\begin{equation}
\frac{M}{Q} = \left(\frac{M}{Q}\right)_0 \left(1+\alpha_1(\tQ) c_1
+\alpha_2(\tQ) c_2+\ldots\right) \,, \quad \frac{\eta}{s} =
\frac{1}{4 \pi}\left(1+\beta_1(\tQ) c_1 +\beta_2(\tQ)
c_2+\ldots\right) \,,
\end{equation}
where the functions $\alpha_i(\tQ), \beta_i(\tQ)$ may read off from
\eqref{mq}, (\ref{theratio}) and near extremality satisfy
$\alpha_i(\tQ)>0, \beta_i(\tQ)<0$. In contrast to the
four-derivative theories considered above, here we observe that
$\eta/s \neq 1/4\pi$ at the extremal limit. Note also that even when
the curvature squared terms are also present, the sign of the first
correction to $\eta/s$ at extremality is determined by the
curvature-cubed terms.

First suppose $c_1<0$.  Then both bounds are satisfied for any $c_2
<0$.  For $c_2>0$, both bounds can hold simultaneously if $c_2 <
\mathrm{min}(|\alpha_1 c_1/\alpha_2|, |\beta_1 c_1/\beta_2|)$, where
$ \mathrm{min}(a,b)$ ($ \mathrm{max}(a,b)$) denotes the smaller
(larger) of $a,b$.

Now suppose $c_1 >0$.  Then the WGC implies that we must have
$c_2<0$.  Both bounds can be satisfied if $|c_2| >
\mathrm{max}(|\alpha_1 c_1/\alpha_2|, |\beta_1 c_1/\beta_2|)$.

The regions in the $(c_1, c_2)$ plane where both bounds are
satisfied are plotted for several values of $\tQ$ in Figure
\ref{c1c2plot}.  For these theories, the plots do not change
substantially if the viscosity bound is relaxed to $4/25\pi$. For
the case of black branes with spherical horizons, see appendix
\ref{sphere}.

\section{Conclusions}
\label{conclusions}

In this work, we examined the WGC and KSS bound for four-derivative
and six-derivative corrections to charged $AdS_5$ black branes. The
WGC states that higher derivative corrections decrease the mass to
charge ratio of extremal black holes. The KSS bound is a lower bound
on the shear viscosity to entropy density ratio, $\eta/s \geq
1/4\pi$. In particular, we studied the interplay of these
constraints with leading and next-to-leading corrections for a
family of toy-models~(\ref{action}). Such constraints on effective
theories might help to distinguish which theories can be UV
completed. First, we calculated the higher derivative corrections to
the mass density in an AdS background using the covariant ADT
method. For the same type of branes, we then calculated corrections
to the shear viscosity (using holographic methods) and the entropy
density (using Wald's formula). Using these results, we analyzed the
constraints on the mass to charge ratio (WGC) and the shear
viscosity to entropy density ratio (KSS bound). This analysis of the
two constraints required comparison of various thermodynamical
quantities and their ratios (energy, charge/chemical potential,
viscosity and entropy) between uncorrected and higher-derivative
corrected branes. In order to make a meaningful comparison, it is
important to consider quantities at the same temperature (e.g. the
canonical ensemble), especially when we compare quantities in
different theories. For this purpose, we rewrote the ratios $M/Q$
and $\eta/s$ using a parameter $\bar{r}_0$ (defined in
(\ref{rbar})), which corresponds to keeping the temperature
unchanged.

One of our main conclusions is that six-derivative corrections in
general behave differently than four-derivative corrections. As
noted in~\cite{Cremonini:2009ih}, for typical examples of
four-derivative corrections (e.g. Gauss-Bonnet) the sign of the
correction to $M/Q$ is opposite relative to the sign of the
correction to $\eta/s$. Thus, the constraints are mutually
exclusive, which might suggest that the theory cannot be UV
completed. In contrast, if we consider a particular coefficient of
the six-derivative corrections (say $c_1$ or $c_2$
in~(\ref{action})), we find that both constraints require the same
sign of the coefficient. Hence, in the sense of the WGC and KSS
bound, six-derivative terms tend to be ``good'' corrections. It is
interesting to note that the ``good'' behavior of the six-derivative
terms might be related to the fact that for those terms the shear
viscosity to entropy density ratio at extremality does not vanish. A
similar behavior was observed for the
$R_{\mu\nu\rho\sigma}\,F^{\mu\nu}\,F^{\rho\sigma}$ term in
four-derivative actions~\cite{Pal}. Note that this ``good'' behavior
does not depend on the exact value of the lower bound on $\eta/s$.
We also found that working to order $\alpha'^2$ in theories with
four-derivative corrections (e.g. Gauss-Bonnet or Weyl-squared)
introduced a wider possibility in the parameter space to satisfy the
constraints.

In the context of this work, we found that the ADT method is a
convenient way to calculate the energy.  The procedure is based on
linearizing the equations of motion and does not require explicit
expressions for boundary counterterms in the action (note that we
also avoided counterterms in the viscosity calculation
following~\cite{MyersChemical}). By construction, the background AdS
space automatically has vanishing energy.  As seen above, the
effective stress tensor \eqref{stress} for the theory \eqref{action}
maintained precisely the same form as in four-derivative gravity. In
particular, the six-derivative terms only changed the coefficients
$\alpha_i$. It would be interesting to understand if this result
holds more generally, say for other curvature-cubed terms or even
higher derivative terms.  One might also try to determine if the
form of \eqref{stress} is in fact the unique tensor with the desired
properties.  If so, it may be possible to find an even faster
general method to extract the corrections to the coefficients
directly from the action.



The action (\ref{action}) contains two six-derivative corrections
that (up to field redefinitions) represent all possible
six-derivative terms when a gauge field is not included.  When a
gauge field is added, there are many more types of possible
correction terms~\cite{MyersChemical}. We considered only a limited
family of corrections in order to learn about the possibilities
opened by including higher than quadratic derivative terms. As a
future direction, it would be interesting to consider the most
general action with six-derivative corrections including a gauge
field. Note that the effective stress-tensor~(\ref{stress})
computation does not depend on the matter part of the action and
therefore is already given in the most general form for
six-derivative corrections.

Finally, our expressions for $M/Q$, $\eta/s$ were given as
perturbative expansions in the couplings $b_i,c_i$, but it would
also be interesting to rewrite these results in terms of physical
CFT parameters, like the central charges $c,a$ or the flux
coefficient $t_4$.  One could then add to the analysis the
constraints on these parameters that arise from various CFT
considerations, such as positivity of the energy
flux~\cite{Hofman:2008ar,Hofman:2009ug}.

\section*{Acknowledgments}
We thank Hari Kunduri for discussions and collaboration during the
initial stages of this work. We are grateful to Ramy Brustein and
Amit Giveon for discussions, as well as Geoffrey Comp\`ere for
correspondence.  We also thank Rob Myers for comments on a draft of
this work. AA acknowledges financial support from the Killam Trust
and partial support from NSERC. DG thanks NSERC for the financial
support.


\appendix

\section{The Linearized Equation of Motion}
\label{lin}

In this appendix we collect some results that are useful for
obtaining the final expression for the energy \eqref{energy}.

To find the effective stress-energy tensor, we wish to linearize the
equations of motion \eqref{eom} about a pure AdS background
satisfying
\begin{equation}
\bar R_{\mu \rho \nu \sigma}= k\, (\bg_{\mu \nu} \bg_{\rho
\sigma}-\bg_{\mu \sigma} \bg_{\rho \nu})\,, \quad \bar R_{\mu
\nu}=(d-1)k\bg_{\mu \nu}\,, \quad \bar R = d(d-1)k \,,
\end{equation}
where
\begin{equation}
k\equiv\frac{2 \Lambda_{eff}}{(d-1)(d-2)} \,.
\end{equation}
Here the cosmological constant $\Lambda_{eff}$ satisfies
\eqref{lambda}, whose perturbative solution is
\begin{eqnarray}
\label{lambdapert}
\Lambda_{eff} &=& \Lambda -\frac{2 (d-4) \Lambda^2 \,(d b_1+b_3)}{(d-2)^2}-\frac{2(d-4)(d-3) \Lambda^2 \,b_2 }{(d-1)(d-2)}-\frac{16 (d-6) \Lambda^3 \,c_1}{(d-2)^3(d-1)^2} \nonumber \\
&&-\frac{4(d-6) \Lambda^3 \,c_2}{(d-1)^2(d-2)^2}+\frac{8 d^2 (d-4)^2  \Lambda^3 \, b_1^2}{(d-2)^4}+\frac{16 d (d-4)^2 \Lambda^3 \, b_1 b_3 }{(d-2)^4} +\frac{8(d-4)^2 \Lambda^3 \,b_3^2}{(d-2)^4}\nonumber \\
&&+\frac{16 (d-4)^2(d-3) \Lambda^3  \,b_2(d
b_1+b_3)}{(d-2)^3(d-1)}+\frac{8(d-4)^2(d-3)^2 \Lambda^3
\,b_2^2}{(d-1)^2(d-2)^2}+\ldots.
\end{eqnarray}

To perform the relevant linearizations, it is useful to recall that
for a decomposition of the metric as $g_{\mu \nu} =\bg_{\mu
\nu}+h_{\mu \nu}$, the linearized Riemann tensor is
\begin{equation}
\left(R^\rho{}_{\mu \lambda \nu}\right)_L = \bnabla_\lambda
\Gamma^\rho_{\mu \nu}-\bnabla_\nu \Gamma^\rho_{\lambda \mu},
\end{equation}
where
\begin{equation}
\Gamma^\rho_{\mu \nu}=\frac{1}{2} \bg^{\rho \sigma}
\left(\bnabla_\mu h_{\nu \sigma}+\bnabla_\nu h_{\mu
\sigma}-\bnabla_\sigma h_{\mu \nu} \right) \,.
\end{equation}
The linearized Ricci tensor is
\begin{equation}
R^L_{\mu \nu}=\frac{1}{2}(-\bar \Box h_{\mu \nu} +\bnabla^\lambda
\bnabla_\nu h_{\mu \lambda}+\bnabla^\lambda \bnabla_\mu h_{\nu
\lambda}-\bnabla_\mu \bnabla_\nu h)
\end{equation}
and the linearized Ricci scalar is
\begin{equation}
R_L = (g^{\mu \nu} R_{\mu \nu})_L=-\bar \Box h +\bnabla^\mu
\bnabla^\nu h_{\mu \nu}-\bar R^{\mu \nu} h_{\mu \nu}\,.
\end{equation}
It is then straightforward (but lengthy) to show that the
linearizations of the six-derivative terms in \eqref{eom} are
\begin{eqnarray}
\left(\nabla_\kappa \nabla_\rho (R_{\mu}{}^{\rho \lambda \tau} R_{\lambda \tau}{}^\kappa{}_\nu) \right)_L &=& -4 k (\bar{\Box}R^{L}_{\mu \nu}-\frac{1}{2} \bnabla_\mu \bnabla_\nu R_L)+4 d k^2 R^{L}_{\mu \nu}-4k^2 R_L \bg_{\mu \nu}\nonumber \\
&&+4k^2(d-1)\bar{\Box} h_{\mu \nu}-4 k^3 d(d-1)h_{\mu \nu}\\
\left( R_{\sigma \mu \lambda \rho}R^{\lambda \rho \alpha \beta} R^{\sigma}{}_{\nu \alpha \beta}\right)_L &=& 12 k^2 R^{L}_{\mu \nu}-8 k^3 (d-1) h_{\mu \nu}\\
\left( \nabla_\kappa \nabla_\rho(R_{\mu \tau \lambda \nu} R^{\rho \tau \lambda \kappa})\right)_L &=& -k \bar{\Box}(R^{L}_{\mu \nu}-\frac{1}{2} \bg_{\mu \nu} R_L)+ k \bnabla_\mu \bnabla_\nu R_L+2 d k^2 R^{L}_{\mu \nu}-2k^2 R_L \bg_{\mu \nu}\nonumber \\
&&+(d-1)k^2\bar{\square} h_{\mu \nu}-2 k^3 d(d-1)h_{\mu \nu}\\
\left( \nabla_\kappa \nabla_\rho(R_\mu{}^{\tau \lambda \kappa} R^\rho{}_{\tau \lambda \nu})\right)_L &=& k \bnabla_\mu \bnabla_\nu R_L+d k^2 R^{L}_{\mu \nu}-k^2 R_L \bg_{\mu \nu}-k^3 d(d-1)h_{\mu \nu} \\
\left(R^\beta{}_{\mu \rho \sigma} R^{\rho \tau}{}_{\lambda \beta}
R^{\sigma}{}_\tau{}^\lambda{}_\nu \right)_L &=& 2(d-3) k^2
R^{L}_{\mu \nu}+k^2 \bg_{\mu \nu} R_L+k^3 (d-1)(4-d) h_{\mu \nu} \,.
\end{eqnarray}
Combining these results with those of \cite{Deser:2002jk} yields
\eqref{stress}.


\section{The Corrected Metric}
\label{CorrectedSolution} We substitute the ansatz (\ref{ansatz1})
in the action (\ref{action}) and obtain 3 equations of motion by
variation with respect to  $\omega(r)$, $\sigma(r)$ and $\gamma(r)$.
Using the zeroth order solution, we can solve the equations of
motion perturbatively in $\alpha'$.  It is convenient to work in a
scheme where the horizon radius is fixed at $r=r_0$, i.e. not
corrected by the higher derivative contributions. This can be
achieved by choosing the integration constants appropriately when
solving the gravitational field equations at each order in
$\alpha'$.   The cosmological constant gets corrected as given in
(\ref{lambdapert}).  We also choose integration constants when
solving Maxwell's equations so that the charge is not corrected and
remains as in the zeroth order. The solution is

\bea
\omega&=&\omega_{(0)}+\omega_{(1)}+\omega_{(2)}\\
\non
\omega_{(1)}&=&\frac{b_1}{6\,r^{10}\,r_0^8}\left[40\,r^8\,r_0^8\(r^4-r_0^4\)
-16\,q^2\,r^2\,r_0^6\(13\,r^6-15\,r^4\,r_0^2+2\,r_0^6\)+q^4\(47\,r_0^8-
32\,r^2\,r_0^6-15\,r^8\)\right]\\ \non &&
+\frac{2\,b_2}{r^{10}r_0^4}\left[q^2\,r^2-(q^2+r^6)r_0^2+r^2\,r_0^6\right]^2
+\frac{b_3}{6\,r^{10}\,r_0^8}\left[8\,r^8\,r_0^8\(r^4-r_0^4\)+
8\,q^2\,r^2r_0^6\(2\,r^6-3\,r^4\,r_0^2+r_0^6\)\right.\\&&\left. \non
+q^4\(13\,r_0^8+8\,r^2\,r_0^6-21\,r^8\)\right] \eea

\bea
\omega_{(2)}&=&-\frac{8\,b_2^2}{r^{16}\,r_0^6}\left[q^2\,r^2-(q^2+r^6)r_0^2+r^2\,r_0^6\right]^3
-\frac{4\,b_1^2}{9\,r^{16}\,r_0^{14}}\left[200\,r^{14}\,r_0^{14}\(r_0^4-r^4\)+24\,q^2\,r^4\,r_0^{12}\(209\,r^{10}\right.\right.
\\ \non && \left.\left.
-175\,r^{8}\,r_0^2+62\,r^4\,r_0^6-96\,r_0^{10}\)\right.+q^4\,r^2\,r_0^6\(6544\,r_0^{12}-4608\,r^2\,r_0^{10}-2773\,r^4\,r_0^8+1824\,r^6\,r_0^6-987\,r^{12}\)\\
\non &&\left.
+q^6\(398\,r^{14}+30\,r^8\,r_0^6-2304\,r^4\,r_0^{10}+6544\,r^2\,r_0^{12}-4668\,r_0^{14}\)\right]
+\frac{2\,b_3^2}{63\,r^{16}\,r_0^{14}}\left[112\,r^{14}\,r_0^{14}\(r^4-r_0^4\)\right.
\\ \non && \left.
+168\,q^2\,r^4\,r_0^{12}\(26\,r^{10}+2\,r^{8}\,r_0^2-67\,r^4\,r_0^6+39\,r_0^{10}\)\right.
+q^6\(7977\,r_0^{14}-1264\,r^{14}+147\,r^8\,r_0^6+6552\,r^4\,r_0^{10}\right.\\
\non &&\left. \left.
-13412\,r^2\,r_0^{12}\)-28\,q^4\,r^2\,r_0^6\(479\,r_0^{12}-468\,r^2\,r_0^{10}-461\,r^4\,r_0^8+408\,r^6\,r_0^6+42\,r^{12}\)\right]\\
\non &&
-\frac{2\,b_1\,b_3}{63\,r^{16}\,r_0^{14}}\left[\right.1120\,r^{14}\,r_0^{14}\(r_0^4-r^4\)
+840\,q^2\,r^4\,r_0^{12}\(8\,r^{10}+11\,r^8\,r_0^2+5\,r^4\,r_0^6-24\,r_0^{10}\)-14\,q^4\,r^2\,r_0^6\,\(69\,r^{12}\right.\\
\non &&
\left.\left.-168\,r^6\,r_0^6+1309\,r^4\,r_0^8+2880\,r^2\,r_0^{10}-4090\,r_0^{12}\)
 +q^6\(4972\,r^{14}+483\,r^8\,r_0^6-20160\,r^4\,r_0^{10}+57260\,r^2\,r_0^{12}\right. \right.
\\ \non && \left.\left. -42555\,r_0^{14}\)\right]
+\frac{2\,b_2\,b_3}{21\,r^{16}\,r_0^{14}}\left[\right.112\,r^{10}r_0^{14}\(r^8+3\,r^4\,r_0^4-4\,r_0^8\)
-28\,q^2\,r^4\,r_0^{12}\(54\,r^{10}+13\,r^8\,r_0^2+38\,r^6\,r_0^4\right.\\
\non && \left.-96\,r^4\,r_0^6-9\,r_0^{10}\)
+7\,q^4\,r^2\,r_0^6\,\(69\,r^{12}-67\,r^8\,r_0^4+408\,r^6\,r_0^6-289\,r^4\,r_0^8+72\,r^2\,r_0^{10}-193\,r_0^{12}\)
\\ \non && \left. +q^6\(1207\,r_0^{14}-1351\,r^2\,r_0^{12}+252\,r^4\,r_0^{10}-147\,r^8\,r_0^6
+147\,r^{10}\,r_0^4-108\,r^{14}\)\right]\\ \non &&
+\frac{2\,b_1\,b_2}{15\,r^{16}\,r_0^{14}}\left[\right.
8\,q^2\,r^4\,\(r_0^2-r^2\)r_0^{12}\(567\,r^8+167\,r^6\,r_0^2-138\,r^4\,r_0^4+219\,r^2\,r_0^6+399\,r_0^8\)+q^4\,r^2\,r_0^6\,\(843\,r^{12} \right.\\
\non && \left.
+1715\,r^8\,r_0^4-3696\,r^6\,r_0^6+785\,r^4\,r_0^8+6384\,r^2\,r_0^{10}-6031\,r_0^{12}\)
 +q^6\(3475\,r_0^{14}-6031\,r^2\,r_0^{12}+3192\,r^4\,r_0^{10}\right.\\ \non &&\left. -480\,r^6\,r_0^8-75\,r^8\,r_0^6
+75\,r^{10}\,r_0^4-156\,r^{14}\) \left.
+80\,r^6\,r_0^{14}\(5\,r^{12}-9\,r^8\,r_0^4+10\,r^4\,r_0^8-6\,r_0^{12}\)\right]
\\
\non && +\frac{c_1}{105\,r^{16}\,r_0^{14}}\left[\right.
84\,q^2\,r^4\,\(r^2-r_0^2\)r_0^{12}\(554\,r^8+959\,r^6\,r_0^2+1289\,r^4\,r_0^4-117\,r^2\,r_0^6-147\,r_0^8\)
+140\,r^6\,r_0^{14}\(r^{12} \right. \\
\non && \left. -94\,r^8\,r_0^4+99\,r^4\,r_0^8-6\,r_0^{12}\)
-21\,q^4\,r^2\,r_0^6\,\(4853\,r^{12}-660\,r^8\,r_0^4+5624\,r^6\,r_0^6-9455\,r^4\,r_0^8-1176\,r^2\,r_0^{10}+814\,r_0^{12}\)
\\ \non && \left. +2\,q^6\(17813\,r^{14}-420\,r^6\,r_0^8+6174\,r^4\,r_0^{10}-8547\,r^2\,r_0^{12}-15020\,r_0^{14}\)\right]
 +\frac{c_2}{140\,r^{16}\,r_0^{14}}\left[\right.
84\,q^2\,r^4\,\(r^2-r_0^2\)r_0^{12}\(112\,r^8 \right. \\
\non &&
\left.+17\,r^6\,r_0^2+107\,r^4\,r_0^4-91\,r^2\,r_0^6-171\,r_0^8\)
+140\,r^6\,r_0^{14}\(r^{12}-12\,r^8\,r_0^4+27\,r^4\,r_0^8-16\,r_0^{12}\)
-63\,q^4\,r^2\,r_0^6\,\(53\,r^{12} \right.\\
\non &&  \left.
-60\,r^8\,r_0^4+264\,r^6\,r_0^6-215\,r^4\,r_0^8-456\,r^2\,r_0^{10}+414\,r_0^{12}\)
 +2\,q^6\(279\,r^{14}-1120\,r^6\,r_0^8+7182\,r^4\,r_0^{10}-13041\,r^2\,r_0^{12} \right.
\\ \non && \left. \left. +6700\,r_0^{14}\)\right]
\eea

\bea
\sigma&=&1+\sigma_{(1)}+\sigma_{(2)}\\
\sigma_{(1)}&=&\frac{4\,q^2}{3\,r^6}\(7\,b_1+5\,b_3\)  \\
\sigma_{(2)}&=&-\frac{8\,b_1^2\,q^2}{9\,r^{12}\,r_0^2}\(3168\,\(q^2+r_0^6\)\,r^2-7\(719\,q^2+176\,r^6\)\,r_0^2\)
-\frac{16\,b_3^2\,q^2}{9\,r^{12}\,r_0^2}\(18\,\(q^2+r_0^6\)\,r^2-70\,r^6\,r_0^2-103\,q^2\,r_0^2\) \non\\
&&+\frac{8\,b_2\,b_3}{3\,r^{12}\,r_0^4}\(30\,r^4\,r_0^{12}+20\,q^2\,r^2\,r_0^4\,\(r^4+3\,r^2\,r_0^2
-9\,r_0^4\)+q^4\,\(30\,r^4-180\,r^2\,r_0^2+193\,r_0^4\)\)\\
&&\non-\frac{8\,b_1\,b_3\,q^2}{9\,r^{12}\,r_0^2}\(1980\,r^2\,r_0^6-1076\,r^6\,r_0^2+5\,q^2\(396\,r^2-673\,r_0^2\)\)\\
&& \non
+\frac{8\,b_2\,b_1}{15\,r^{12}\,r_0^4}\(270\,r^4\,r_0^{12}+q^4\(270\,r^4-1692\,r^2\,r_0^2
+1855\,r_0^4\)+4\,q^2\(35\,r^6\,r_0^4+135\,r^4\,r_0^6-423\,r^2\,r_0^8\)\)
\\ && \non
-\frac{2\,c_1}{5\,r^{12}\,r_0^4}\(390\,r^4\,r_0^{12}+q^4\(390\,r^4-3564\,r^2\,r_0^2+7705\,r_0^4\)
+4\,q^2\(65\,r^6\,r_0^4+195\,r^4\,r_0^6-891\,r^2\,r_0^8\)\)
\\ && \non
+\frac{3\,c_2}{10\,r^{12}\,r_0^4}\(30\,r^4\,r_0^{12}+q^4\(30\,r^4-108\,r^2\,r_0^2+85\,r_0^4\)
+4\,q^2\(5\,r^6\,r_0^4+15\,r^4\,r_0^6-27\,r^2\,r_0^8\)\) \eea

\bea
\gamma&=&\gamma_{(0)}+\gamma_{(1)}+\gamma_{(2)}\\
\frac{\gamma_{(1)}}{\gamma_{(0)}}&=&\frac{q^2}{3\,r^6}\(7\,b_1+5\,b_3\)\\
\frac{\gamma_{(2)}}{\gamma_{(0)}}&=&-\frac{8\,b_1^2\,q^2}{9\,r^{12}r_0^2}\(528\,r^2\,(r_0^6+q^2)-308\,r^6\,r_0^2-719\,q^2\,r_0^2\)
-\frac{8\,b_3^2\,q^2}{9\,r^{12}r_0^2}\(42\,r^2\,(r_0^6+q^2)-245\,r^6\,r_0^2-206\,q^2\,r_0^2\)\\
\non &&
-\frac{8\,b_1\,b_3\,q^2}{63\,r^{12}\,r_0^2}\(2310\,r^2\,r_0^6-1883\,r^6\,r_0^2+5\,q^2\,\(462\,r^2-673\,r_0^2\)\)
\\ \non && +\frac{8\,b_1\,b_2}{15\,r^{12}\,r_0^4}\(54\,r^4\,\(r_0^{12}+q^4\)+q^4\,r_0^2\(265\,r_0^2-282\,r^2\)
+q^2\,r_0^4\(35\,r^6+108\,r^4\,r_0^2-282\,r^2\,r_0^4\)\) \\ \non &&
+\frac{8\,b_2\,b_3}{21\,r^{12}\,r_0^4}\(42\,r^4\(r_0^{12}+q^4\)+q^4\,r_0^2\(193\,r_0^2-210\,r^2\)
+7\,q^2\,r_0^4\(5\,r^6+12\,r^4\,r_0^2-30\,r^2\,r_0^4\)\) \\ \non &&
+\frac{2\,c_1}{35\,r^{12}\,r_0^4}\(546\,r^4\(r_0^{12}+q^4\)+q^4\,r_0^2\(7705\,r_0^2-4158\,r^2\)
+7\,q^2\,r_0^4\(65\,r^6+156\,r^4\,r_0^2-594\,r^2\,r_0^4\)\) \\ \non
&&
+\frac{3\,c_2}{70\,r^{12}\,r_0^4}\(42\,r^4\(r_0^{12}+q^4\)+q^4\,r_0^2\(85\,r_0^2-126\,r^2\)
+7\,q^2\,r_0^4\(5\,r^6+12\,r^4\,r_0^2-18\,r^2\,r_0^4\)\). \eea

Substituting in eq. (\ref{temperature}), we get the temperature of
the corrected solution as: \bea \non
T&=&\frac{r_0}{\pi}\left[1-\frac{\tilde{q}^2}{2}-\frac{1}{3}
\left(b_1 \left(9 \tilde{q}^4+64 \tilde{q}^2-20\right)+b_3 \left(9
\tilde{q}^4+20
   \tilde{q}^2-4\right)\right)+\frac{c_1}{15} \left(4583 \tilde{q}^6-138 \tilde{q}^4-1086 \tilde{q}^2+620\right)\right.\\
  &&+\frac{3\,c_2}{20}  \left(\tilde{q}^2-2\right)\left(3 \tilde{q}^4-12 \tilde{q}^2+10\right)-\frac{4\,b_2}{15} \left(\tilde{q}^2-2\right) \left(5\, b_3 \left(9 \tilde{q}^4
  -12
   \tilde{q}^2-10\right)+b_1\left(93 \tilde{q}^4-132 \tilde{q}^2-130\right)\right)\non\\&&
  -\frac{4\,b_1^2}{9} \left(493 \tilde{q}^6-84 \tilde{q}^4+1896
   \tilde{q}^2-200\right)
 -\frac{4\,b_3^2}{9}\left(91 \tilde{q}^6+222 \tilde{q}^4+204 \non
   \tilde{q}^2-8\right)\\&&\left.
   -\frac{8\,b_1\,b_3}{9} \left(212 \tilde{q}^6+255 \tilde{q}^4+570
   \tilde{q}^2-40\right)\right],
   \eea
where $\tilde{q}\equiv q/r_0^3$ is a dimensionless charge parameter.
We write the CFT temperature $T_{CFT}\equiv \ell_{eff} T$ so that
the higher derivative corrections are absorbed in a redefined
parameter $\bar{r}_0$:

\bea r_0&=&\br\left[1+\frac{2}{15 \bar{q}^2+6}\(b_1\(9 \bar{q}^4+59
\bar{q}^2-10\)-3\,b_2\(\bar{q}^2-2\)+b_3\(9 \bar{q}^4+19
\bar{q}^2-2\)\) \right. \non\\\non &&+\frac{2\,b_2\,
\left(\bar{q}^2-2\right) }{5 \left(5\, \bar{q}^2+2\right)^3}
\left(5\,b_3 \left(300 \bar{q}^8-85 \bar{q}^6-621 \bar{q}^4-284
\bar{q}^2-60\) \left.+b_1\left(3100 \bar{q}^8-1545 \bar{q}^6-8769
\bar{q}^4-3604 \bar{q}^2-860\right)\right)\) \label{rbar2}\\&&\left.
+\frac{4\,b_1\, b_3}{9
   \left(5 \bar{q}^2+2\right)^3} \left(14720 \bar{q}^{10}+2446 \bar{q}^8+32263 \bar{q}^6+20594 \bar{q}^4
   +10228 \bar{q}^2-200\right) \right.\\
   \non &&+\frac{2\,b_1^2}{9
   \left(5 \bar{q}^2+2\right)^3} \left(42820 \bar{q}^{10}-27874\bar{q}^8+69839\bar{q}^6+48106 \bar{q}^4+33716\bar{q}^2-1000\right)\\ &&\left. +\frac{2\,b_3^2}{9
   \left(5 \bar{q}^2+2\right)^3}\left(2620\bar{q}^{10}+8366\bar{q}^8+20687{q}^6+11290\bar{q}^4+3572 \bar{q}^2-40\right)\right. \non \\
  \non && \left.-\frac{2\,  c_1}{15 \left(5 \bar{q}^2+2\right)} \left(4583 \bar{q}^6-138 \bar{q}^4-1081
  \bar{q}^2+610\right)-\frac{c_2}{10\(\bar{q}^2+2\)}
  \(q^2-2\)\(3\,q^2-7\)\(3\,q^2-5\)
   \right],
\eea where we have set $\tQ \equiv q/\bar r_0^{\,3}$.

\section{Mass to Charge Ratio for Spherical Horizons}
\label{sphere}

In the text above, we discussed $M/Q$ for planar black hole
solutions.  However, given that the original formulation of the WGC~\cite{weak1}
 applied to asymptotically flat extremal black holes, it
is perhaps more appropriate to consider $M/Q$ for black holes with a
spherical horizon~\cite{MyersChemical}.  This further allows us to check another aspect
of the WGC, namely the prediction that the correction to the mass to
charge ratio should become more negative for smaller extremal black
holes.

It is straightforward to repeat the calculations of section
\ref{mq5d} for the case of spherical horizons, i.e. with a metric
ansatz
 \begin{eqnarray}
ds^2 =-\omega(r) dt^2+\frac{\sigma^2(r)}{\omega(r)}dr^2+r^2 d\Omega^2_3\,, \\
\end{eqnarray}
where $d\Omega^2_3$ is the line element of the unit $S^3$.  In the
absence of higher derivative corrections, the solution to the
equations of motion is now given by
\begin{eqnarray}
\omega_{(0)}(r) &=& 1+r^2-\frac{q^2+ r_0^4+r_0^6}{r_0^2 \, r^2 }+\frac{q^2}{r^4} \\
\sigma_{(0)}(r) &=& 1\\
\gamma_{(0)}(r) &=& \frac{\sqrt{3} \,q }{r^2} \,.
\end{eqnarray}
The horizon is at $r=r_0$ and the extremal case is $q^2 = 2 r_0^6
(1+\frac{1}{2r_0^2})$.

In the interest of brevity, we omit the full solution and the mass
to charge ratio for general $q$ when the higher derivative
corrections are included.  Instead we just give the result for the
mass to charge ratio in the extremal case:
\begin{eqnarray}
\label{mqsphere} \frac{M}{Q} &=& \left(\frac{M}{Q}\right)_0
\left(1-\frac{264 \bar{r}_0^4+284 \bar{r}_0^2+1}{6 \bar{r}_0^2
\left(3\bar{r}_0^2+2\right)}b_1 -\frac{3 \bar{r}_0^4+2
\bar{r}_0^2-2}{\bar{r}_0^2 \left(3\bar{r}_0^2+2\right)}b_2
-\frac{\left(8 \bar{r}_0^2+1\right) \left(12
\bar{r}_0^2+11\right)}{6 \bar{r}_0^2 \left(3
   \bar{r}_0^2+2\right)}b_3
\right. \nonumber \\
&&+\frac{31182 \bar{r}_0^6+29392
\bar{r}_0^4+8865\bar{r}_0^2+536}{105 \bar{r}_0^4 \left(3
\bar{r}_0^2+2\right)}c_1 +\frac{ 366 \bar{r}_0^6+136 \bar{r}_0^4-765
\bar{r}_0^2-152}{140\bar{r}_0^4 \left(3 \bar{r}_0^2+2\right)}c_2
-\frac{15 \bar{r}_0^4+10 \bar{r}_0^2+4}{2\bar{r}_0^2 \left(3 \bar{r}_0^2+2\right)}b_2^2\nonumber \\
&&-\frac{103950 \bar{r}_0^8+99306 \bar{r}_0^6+23013 \bar{r}_0^4-929
\bar{r}_0^2-67}{45 \bar{r}_0^4\left(3 \bar{r}_0^2+1\right) \left(3
\bar{r}_0^2+2\right)}b_1^2 -\frac{1752 \bar{r}_0^6+52
\bar{r}_0^4-541
   \bar{r}_0^2-96}{42 \bar{r}_0^4 \left(3 \bar{r}_0^2+2\right)}b_2 b_3
\nonumber \\
&&-\frac{19170
\bar{r}_0^8+19710\bar{r}_0^6+5661\bar{r}_0^4+383\bar{r}_0^2+55}{63\bar{r}_0^4\left(3\bar{r}_0^2
+1\right)\left(3\bar{r}_0^2+2\right)}b_3^2 -\frac{4128
\bar{r}_0^6+1348
   \bar{r}_0^4-325 \bar{r}_0^2-96}{30 \bar{r}_0^4 \left(3 \bar{r}_0^2+2\right)}b_2 b_1
\nonumber \\
&&\left.-\frac{2  \left(50814
   \bar{r}_0^8+48870 \bar{r}_0^6+11322 \bar{r}_0^4-326 \bar{r}_0^2+5\right)}{63 \bar{r}_0^4 \left(3
   \bar{r}_0^2+1\right) \left(3 \bar{r}_0^2+2\right)}b_3 b_1+ \ldots \right)
\end{eqnarray}
where
\begin{equation}
\left(\frac{M}{Q}\right)_0 = \frac{\sqrt{3} \left(3
\bar{r}_0^2+2\right)}{64\,\pi\,G \sqrt{2 \bar{r}_0^2+1}} \,.
\end{equation}
Similarly to section \ref{mq5d}, the parameter $\bar r_0$ has been
defined by fixing
\begin{equation}
T_{CFT} = \frac{\bar r_0}{\pi}\left(1-\frac{q^2}{2\bar
r_0^{\,6}}+\frac{1}{2\bar r_0^2}\right) \,,
\end{equation}
with the extremal case corresponding to $q^2 = 2 \bar
r_0^6\left(1+\frac{1}{2\bar r_0^2}\right)$.  This reparametrization
ensures that the uncorrected and corrected values of $M/Q$ are
compared at the same temperature.  Note that \eqref{mqsphere}
reduces to the planar black hole result \eqref{mqex} in the limit
$\bar{r}_0\to\infty$.

In the case of spherical horizons, the analysis of the two
constraints is qualitatively similar to the planar case until $\bar
r_0 \lesssim 1$, at which point the behavior of $M/Q$ may change.
However, it is plausible that the exact black hole solution might
not exist for $\bar{r}_0$ of order the AdS radius.  Let us consider
for example theories in which the first corrections to the effective
action involve six-derivatives, i.e., $b_1 = b_2= b_3 = 0$.

First suppose $c_2=0$.  The coefficient of $c_1$ in \eqref{mqsphere}
is positive for all $\bar r_0$, so the WGC requires $c_1<0$. Note
that because of the factor $\bar r_0^4$ in the denominator, the
correction becomes more negative for smaller $\bar r_0$ as expected.
When $c_1<0$, the KSS bound is also satisfied.

Now suppose $c_1=0$.  The coefficient of $c_2$ in \eqref{mqsphere}
is positive for large $\bar r_0$, but becomes negative for $\bar r_0
\lesssim 1.17$. The KSS bound meanwhile requires $c_2 <0$, so both
bounds can be satisfied only when $\bar r_0 \gtrsim 1.17$ and $c_2
<0$. Once again, the correction to $M/Q$ becomes more negative for
smaller $\bar r_0$.

\begin{figure}[h]
\begin{picture}(0,0)
\put(-124,-40){$\bar r_0 = 10$} \put(24,-40){$\bar r_0 = 1$}
\put(174,-40){$\bar r_0 = 0.1$} \put(-225,-82){$c_2$}
\put(-142,-160){$c_1$}
\end{picture}
\begin{center}
\includegraphics[width=6in]{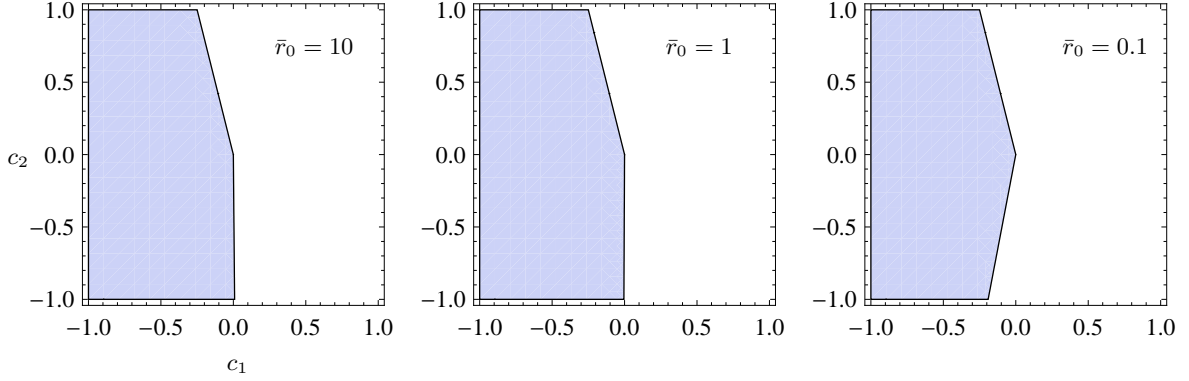}
\end{center}
\caption{Here we consider an extremal black hole with a spherical
horizon in a six-derivative theory with $b_1 = b_2= b_3 = 0$.  The
shaded regions represent points in the $(c_1, c_2)$ plane (for
various values of $\bar r_0$) where both the WGC
and the KSS bound are satisfied simultaneously.}
\label{c1c2sphereplot}
\end{figure}

For the general case, the regions in the $(c_1, c_2)$ plane where both bounds are
satisfied are plotted for several values of $\bar r_0$ in Figure
\ref{c1c2sphereplot}.

\section{Expressions for the Effective Action}
In this appendix we give the coefficients of the effective action
(\ref{effective}) obtained from the action (\ref{action2}), which
are needed for the computation of the viscosity (\ref{visc_final}).
Substituting the ansatz given in (\ref{ansatz1}) yields
\label{expVisco} \bea \nonumber
 A(r)&=&\frac{2\,r^3\,\omega}{\sigma}\left[1
 - \frac{2\,{b_1}\,\left(
      \sigma \,\left( 6\,\omega  + 6\,r\,\omega ' + r^2\,\omega '' \right)- r\,\sigma '\,\left( 6\,\omega  + r\,\omega ' \right)    \right) }{r^2\,{\sigma }^3}-  \frac{4\,{b_2}\,\left( \omega  + r\,\omega ' \right) }{r^2\,{\sigma }^2}\right.
       +\frac{3\,{c_1}\,{\left( \sigma \,\omega ' -2\,\omega
\,\sigma '\right) }^2}
   {r^2\,{\sigma }^6}\\ &&+\frac{{b_3}\,\left( r\,\sigma '\,\left( 8\,\omega  + r\,\omega ' \right)  -
      \sigma \,\left( 4\,\omega  + 5\,r\,\omega ' + r^2\,\omega '' \right)  \right) }{2\,r^2\,{\sigma }^3}
\left. +
  \frac{3\,{c_2}\,\left(
      \sigma \,\omega '\,\left( 4\,\omega  + r^2\,\omega '' \right)- \sigma '\,\left( 8\,{\omega }^2 + r^2\,{\omega '}^2   \right)  \right) }{4\,r^3\,{\sigma }^5}\right]
\\ B(r)&=&\frac{3\,r^3\,\omega }{2\,\sigma }\left[1+\frac{2\,{b_1}\,
\left( r\,\sigma '\,\left( 6\,\omega + r\,\omega ' \right)   -
      \sigma \,\left( 6\,\omega  + 6\,r\,\omega ' + r^2\,\omega '' \right)  \right) }{r^2\,{\sigma }^3}\right.\\
&&- \nonumber \frac{2\,{b_2}\,\left( 12\,\sigma \,{\omega }^2 -
6\,r\,{\omega }^2\,\sigma ' +
      14\,r\,\sigma \,\omega \,\omega ' - 3\,r^2\,\omega \,\sigma '\,\omega ' +
      r^2\,\sigma \,\(\omega '\omega\)'  \right) }{3\,r^2\,
    {\sigma }^3\,\omega }\\&& \nonumber +\frac{{b_3}\,\left( r^2\,{\omega }^2\,{\sigma '}^2 +
      r\,\sigma \,\omega \,\sigma '\,\left( 4\,\omega  - r\,\omega ' \right)  +
      {\sigma }^2\,\left(r^2\,{\omega '}^2  - r\,\omega \,\omega '   +
         \omega \,\left( \omega  - r^2\,\omega '' \right)  \right)  \right) }{3\,r^2\,{\sigma }^4\,\omega
         }\\&&\nonumber+\frac{2\,{c_1}}{r^4\,{\sigma }^7\,\omega }\left(4\,r^3\,{\omega }^3\,{\sigma '}^3 - 6\,r^2\,\sigma \,{\omega }^2\,{\sigma '}^2\,\left( 2\,\omega  + r\,\omega ' \right)  +
  r\,{\sigma }^2\,\omega \,\sigma '\,\left( 16\,{\omega }^2 + 12\,r\,\omega \,\omega ' + 3\,r^2\,{\omega '}^2
  \right)\right.\\&& \nonumber \left.-{\sigma }^3\,\left( 2\,{\omega }^3 + 8\,r\,{\omega }^2\,\omega ' + 3\,r^2\,\omega \,{\omega '}^2 +
      r^3\,{\omega '}^3 \right) \right)
      \\&& \nonumber +\frac{c_2}{4\,r^4\,{\sigma }^6\,\omega}\( r\,\sigma \,\sigma '\,
   \left(   2\,r^2\,\omega \,{\omega '}^2 -16\,{\omega }^3+ r^3\,{\omega '}^3 + 2\,r^3\,\omega \,\omega '\,\omega ''
     \right)-2\,r^4\,\omega \,{\sigma '}^2\,{\omega '}^2\right.
      \\
&& \nonumber \left. \left.+{\sigma }^2\,\left( 24\,{\omega }^3 -
r^4\,{\omega '}^2\,\omega '' -
    2\,r\,\omega \,\omega '\,\left( r^2\,\omega '' -4\,\omega\right)
    \right)\)\right]\eea
    \bea
    E(r)&=&\frac{\omega^2}{2\sigma^3}\left[b_3\,r^3-\frac{12\,c_1\,r^2\(\sigma\omega'-2\omega\sigma'\)}{\sigma^3}+\frac{3\,c_2\,\omega\,r}{\sigma^2}\right]
   \\ F(r)&=&-\frac{2\,\omega}{\sigma^3}\left[r^2\,{b_2}\,\left( 2\,\omega
+ r\,\omega ' \right)  -
  \frac{r^2\,{b_3}\,\left( 3\,\sigma \,\omega  - r\,\omega \,\sigma ' + r\,\sigma \,\omega ' \right) }{2\,\sigma } +
  \frac{3\,r\,{c_1}\,\left(  \sigma \,\omega '-2\,\omega \,\sigma '  \right) \,
     \left( 4\,\sigma \,\omega  - 2\,r\,\omega \,\sigma ' + r\,\sigma \,\omega ' \right) }{{\sigma
     }^4}\right. \nonumber
     \\&&\left.  + \frac{3\,{c_2}\,\left(
       r^3\,\sigma \,\omega '\,\omega '' -8\,\sigma \,{\omega }^2 - r^3\,\sigma '\,{\omega '}^2  \right) }{8\,{\sigma }^3}\right]
    \eea



\begin{thebibliography}{11}

\bibitem{weak1}
  N.~Arkani-Hamed, L.~Motl, A.~Nicolis and C.~Vafa,
  ``The string landscape, black holes and gravity as the weakest force,''
  JHEP {\bf 0706}, 060 (2007)
  [arXiv:hep-th/0601001].

\bibitem{KSS}
  P.~Kovtun, D.~T.~Son and A.~O.~Starinets,
  ``Viscosity in strongly interacting quantum field theories from black hole
  physics,''
  Phys.\ Rev.\ Lett.\  {\bf 94}, 111601 (2005)
  [arXiv:hep-th/0405231].


\bibitem{Vafa}
  C.~Vafa,
  ``The string landscape and the swampland,''
  arXiv:hep-th/0509212.

\bibitem {blmsy} M.~Brigante, H.~Liu, R.~C.~Myers, S.~Shenker and S.~Yaida,
  ``Viscosity Bound Violation in Higher Derivative Gravity,''
  Phys.\ Rev.\  D {\bf 77}, 126006 (2008),
  [arXiv:0712.0805 [hep-th]].

\bibitem{Cremonini:2009ih}
  S.~Cremonini, J.~T.~Liu and P.~Szepietowski,
  ``Higher Derivative Corrections to R-charged Black Holes: Boundary
  Counterterms and the Mass-Charge Relation,''
  JHEP {\bf 1003}, 042 (2010)
  [arXiv:0910.5159 [hep-th]].

\bibitem{Kats2}
  Y.~Kats and P.~Petrov,
  ``Effect of curvature squared corrections in AdS on the viscosity of the dual
  gauge theory,''
  JHEP {\bf 0901}, 044 (2009)
  [arXiv:0712.0743 [hep-th]].

\bibitem{MyersChemical}
R.~C.~Myers, M.~F.~Paulos and A.~Sinha,
  ``Holographic Hydrodynamics with a Chemical Potential,''
  JHEP {\bf 0906}, 006 (2009)
  [arXiv:0903.2834 [hep-th]].



\bibitem{Cremonini}
  S.~Cremonini, K.~Hanaki, J.~T.~Liu and P.~Szepietowski,
  ``Higher derivative effects on eta/s at finite chemical potential,''
  Phys.\ Rev.\  D {\bf 80}, 025002 (2009)
  [arXiv:0903.3244 [hep-th]].

\bibitem{Pal}
  S.~S.~Pal,
  ``Weak Gravity Conjecture, Central Charges and $\eta/s$,''
  arXiv:1003.0745 [hep-th].

\bibitem{weak2}
  Y.~Kats, L.~Motl and M.~Padi,
  ``Higher-order corrections to mass-charge relation of extremal black holes,''
  JHEP {\bf 0712}, 068 (2007)
  [arXiv:hep-th/0606100].

\bibitem{banks}
   T.~Banks, M.~Johnson and A.~Shomer,
  ``A note on gauge theories coupled to gravity,''
  JHEP {\bf 0609}, 049 (2006)
  [arXiv:hep-th/0606277].


\bibitem{weak3}
A.~Giveon and D.~Gorbonos,
  ``On black fundamental strings,''
  JHEP {\bf 0610}, 038 (2006)
  [arXiv:hep-th/0606156].

\bibitem{weak4}
  A.~Giveon, D.~Gorbonos and M.~Stern,
  ``Fundamental Strings and Higher Derivative Corrections to d-Dimensional
  Black Holes,''
  JHEP {\bf 1002}, 012 (2010)
  [arXiv:0909.5264 [hep-th]].


\bibitem{buchel}
A.~Buchel, J.~T.~Liu and A.~O.~Starinets,
  ``Coupling constant dependence of the shear viscosity in N=4 supersymmetric
  Yang-Mills theory,''
  Nucl.\ Phys.\  B {\bf 707}, 56 (2005)
  [arXiv:hep-th/0406264].

\bibitem{buchel2}
A.~Buchel,
  ``Resolving disagreement for eta/s in a CFT plasma at finite coupling,''
  Nucl.\ Phys.\  B {\bf 803}, 166 (2008)
  [arXiv:0805.2683 [hep-th]].

\bibitem{buchel3}
R.~C.~Myers, M.~F.~Paulos and A.~Sinha,
  ``Quantum corrections to eta/s,''
  Phys.\ Rev.\  D {\bf 79}, 041901 (2009)
  [arXiv:0806.2156 [hep-th]].


\bibitem{Ge1} X.~H.~Ge, Y.~Matsuo, F.~W.~Shu, S.~J.~Sin and T.~Tsukioka,
``Viscosity Bound, Causality Violation and Instability with Stringy
Correction and Charge,'' JHEP {\bf 10}, 009 (2008), [arXiv:0808.2354
[hep-th]].

\bibitem{Kout}
G.~Koutsoumbas, E.~ Papantonopoulos, G.~Siopsis, ``Shear Viscosity
and Chern-Simons Diffusion Rate from Hyperbolic Horizons,'' Phys.
Lett. {\bf B 677}, 74 (2009), [arXiv:0809.3388 [hep-th]].

\bibitem{Cai1}
R. G. Cai, Z. Y. Nie and Y. W. Sun, ``Shear Viscosity from Effective
Couplings of Gravitons,'' Phys. Rev. D {\bf 78}, 126007 (2008)
[arXiv:0811.1665 [hep-th]].

\bibitem{bms}
``A. Buchel, R. C. Myers and A. Sinha, Beyond $\eta/s = 1/4\pi$'',
JHEP {\bf 03}, 084 (2009), [arXiv:0812.2521 [hep-th]].

\bibitem{Cai2}
R. G. Cai, Z. Y. Nie, N. Ohta and Y. W. Sun, ``Shear Viscosity from
Gauss-Bonnet Gravity with a Dilaton Coupling,'' Phys. Rev. D {\bf
79}, 066004 (2009) [arXiv:0901.1421 [hep-th]].

\bibitem{ag}
A.~Ghodsi and M.~Alishahiha,
  ``Non-relativistic D3-brane in the presence of higher derivative
  corrections,''
  Phys.\ Rev.\  D {\bf 80}, 026004 (2009)
  [arXiv:0901.3431 [hep-th]].

\bibitem{sm} A.~Sinha and R. C. Myers, ``The viscosity bound in string theory,'' Nucl. Phys. A {\bf  830}, 295C
(2009) [arXiv:0907.4798 [hep-th]].


\bibitem{ssp}S.~S.~Pal,
``$\eta/s$ at finite coupling,'' Phys. Rev. D {\bf 81}, 045005
(2010), [arXiv:0910.0101 [hep-th]].

\bibitem{deBoer}
  J.~de Boer, M.~Kulaxizi and A.~Parnachev,
  ``$AdS_7/CFT_6$, Gauss-Bonnet Gravity, and Viscosity Bound,''
  JHEP {\bf 1003}, 087 (2010)
  [arXiv:0910.5347 [hep-th]].

\bibitem{Camanho}
  X.~O.~Camanho and J.~D.~Edelstein,
  ``Causality constraints in AdS/CFT from conformal collider physics and
  Gauss-Bonnet gravity,''
  JHEP {\bf 1004}, 007 (2010)
  [arXiv:0911.3160 [hep-th]].

\bibitem{Camanho2}
  X.~O.~Camanho and J.~D.~Edelstein,
  ``Causality in AdS/CFT and Lovelock theory,''
  JHEP {\bf 1006}, 099 (2010)
  [arXiv:0912.1944 [hep-th]].











\bibitem{bm} R.~Brustein and A.~J.~M.~Medved, ``Proof of a universal lower
bound on the shear viscosity to entropy density ratio,'' [arXiv:
0908.1473 [hep-th]].


\bibitem{causality}
  M.~Brigante, H.~Liu, R.~C.~Myers, S.~Shenker and S.~Yaida,
  ``The Viscosity Bound and Causality Violation,''
  Phys.\ Rev.\ Lett.\  {\bf 100}, 191601 (2008)
  [arXiv:0802.3318 [hep-th]].

  \bibitem{Dutta2}
  N.~Banerjee and S.~Dutta,
  ``Near-Horizon Analysis of $\eta/s$,''
  arXiv:0911.0557 [hep-th].

\bibitem{Myers1}
  R.~C.~Myers and B.~Robinson,
  ``Black Holes in Quasi-topological Gravity,''
  arXiv:1003.5357 [gr-qc].

\bibitem{Myers2}
  R.~C.~Myers, M.~F.~Paulos and A.~Sinha,
  ``Holographic studies of quasi-topological gravity,''
  arXiv:1004.2055 [hep-th].

\bibitem{Hofman:2008ar}
  D.~M.~Hofman and J.~Maldacena,
  ``Conformal collider physics: Energy and charge correlations,''
  JHEP {\bf 0805}, 012 (2008)
  [arXiv:0803.1467 [hep-th]].


\bibitem{tseytlin}
  R.~R.~Metsaev and A.~A.~Tseytlin,
  ``Curvature Cubed Terms In String Theory Effective Actions,''
  Phys.\ Lett.\  B {\bf 185}, 52 (1987).



\bibitem{Abbott:1981ff}
  L.~F.~Abbott and S.~Deser,
  ``Stability of gravity with a cosmological constant,''
  Nucl.\ Phys.\  B {\bf 195}, 76 (1982).

\bibitem{Deser:2002rt}
  S.~Deser and B.~Tekin,
  ``Gravitational energy in quadratic curvature gravities,''
  Phys.\ Rev.\ Lett.\  {\bf 89}, 101101 (2002)
  [arXiv:hep-th/0205318].

\bibitem{Deser:2002jk}
  S.~Deser and B.~Tekin,
  ``Energy in generic higher curvature gravity theories,''
  Phys.\ Rev.\  D {\bf 67}, 084009 (2003)
  [arXiv:hep-th/0212292].



\bibitem{ADM}
  R.~L.~Arnowitt, S.~Deser, and C.~W.~Misner,
  ``Dynamical structure and definition of energy in general relativity,''
  Phys.\ Rev.\  {\bf 116}, 1322 (1959);

 ``Canonical variables for general relativity,''
  Phys.\ Rev.\  {\bf 117}, 1595 (1960);

  ``The dynamics of general relativity,'' in
{\it Gravitation: an Introduction to Current Research}, L.~Witten
ed. (Wiley 1962), pp 227-265.



\bibitem{LL}
L.~D.~Landau and E.~M.~Lifshitz, {\it The Classical Theory of
Fields}, Pergamon Press, Oxford (1975).

\bibitem{Balasubramanian:1999re}
  V.~Balasubramanian and P.~Kraus,
  ``A stress tensor for anti-de Sitter gravity,''
  Commun.\ Math.\ Phys.\  {\bf 208}, 413 (1999)
  [arXiv:hep-th/9902121].

\bibitem{Henningson:1998gx}
  M.~Henningson and K.~Skenderis,
  ``The holographic Weyl anomaly,''
  JHEP {\bf 9807}, 023 (1998)
  [arXiv:hep-th/9806087].

\bibitem{Hawking:1980gf}
  S.~W.~Hawking,
  ``The path integral approach to quantum gravity,''
  in {\it General Relativity:  An Einstein Centenary Survey}, eds. S.~W.~Hawking and W.~Israel, Cambridge University Press, Cambridge (1979).

\bibitem{Henneaux:1985tv}
  M.~Henneaux and C.~Teitelboim,
  ``Asymptotically Anti-De Sitter Spaces,''
  Commun.\ Math.\ Phys.\  {\bf 98}, 391 (1985).

\bibitem{Barnich:2001jy}
  G.~Barnich and F.~Brandt,
  ``Covariant theory of asymptotic symmetries, conservation laws and  central
  charges,''
  Nucl.\ Phys.\  B {\bf 633}, 3 (2002)
  [arXiv:hep-th/0111246].

\bibitem{Barnich:2007bf}
  G.~Barnich and G.~Comp\`ere,
  ``Surface charge algebra in gauge theories and thermodynamic integrability,''
  J.\ Math.\ Phys.\  {\bf 49}, 042901 (2008)
  [arXiv:0708.2378 [gr-qc]].

\bibitem{Compere:2007az}
  G.~Comp\`ere,
  ``Symmetries and conservation laws in Lagrangian gauge theories with
  applications to the mechanics of black holes and to gravity in three
  dimensions,''
  arXiv:0708.3153 [hep-th].

\bibitem{Lee:1990nz}
  J.~Lee and R.~M.~Wald,
  ``Local symmetries and constraints,''
  J.\ Math.\ Phys.\  {\bf 31}, 725 (1990).

\bibitem{wald2}
 V.~Iyer and R.~M.~Wald,
  ``Some properties of Noether charge and a proposal for dynamical black hole
  entropy,''
  Phys.\ Rev.\  D {\bf 50}, 846 (1994)
  [arXiv:gr-qc/9403028].

\bibitem{Wald:1999wa}
  R.~M.~Wald and A.~Zoupas,
  ``A General Definition of "Conserved Quantities" in General Relativity and
  Other Theories of Gravity,''
  Phys.\ Rev.\  D {\bf 61}, 084027 (2000)
  [arXiv:gr-qc/9911095].

\bibitem{Son1}
  D.~T.~Son and A.~O.~Starinets,
  ``Minkowski-space correlators in AdS/CFT correspondence: Recipe and
  applications,''
  JHEP {\bf 0209}, 042 (2002)
  [arXiv:hep-th/0205051].

\bibitem{Policastro}
  G.~Policastro, D.~T.~Son and A.~O.~Starinets,
  ``From AdS/CFT correspondence to hydrodynamics,''
  JHEP {\bf 0209}, 043 (2002)
  [arXiv:hep-th/0205052].


\bibitem{Dutta1}
N.~Banerjee and S.~Dutta,
  ``Higher Derivative Corrections to Shear Viscosity from Graviton's Effective
  Coupling,''
  JHEP {\bf 0903}, 116 (2009)
  [arXiv:0901.3848 [hep-th]].

\bibitem{liu+iqbal}
N.~Iqbal and H.~Liu,
  ``Universality of the hydrodynamic limit in AdS/CFT and the membrane
  paradigm,'' Phys. Rev. D {\bf  79}, 025023 (2009),
  [arXiv:0809.3808 [hep-th]].

\bibitem{wald1}
  R.~M.~Wald,
  ``Black hole entropy is the Noether charge,''
  Phys.\ Rev.\  D {\bf 48}, 3427 (1993)
  [arXiv:gr-qc/9307038].

\bibitem{myers}
  T.~Jacobson, G.~Kang and R.~C.~Myers,
  ``On Black Hole Entropy,''
  Phys.\ Rev.\  D {\bf 49}, 6587 (1994)
  [arXiv:gr-qc/9312023].


\bibitem{Tseytlin1}
  R.~R.~Metsaev and A.~A.~Tseytlin,
  ``Order alpha-prime (Two Loop) Equivalence of the String Equations of Motion
  and the Sigma Model Weyl Invariance Conditions: Dependence on the Dilaton and
  the Antisymmetric Tensor,''
  Nucl.\ Phys.\  B {\bf 293}, 385 (1987).


\bibitem{Hofman:2009ug}
  D.~M.~Hofman,
  ``Higher Derivative Gravity, Causality and Positivity of Energy in a UV
  complete QFT,''
  Nucl.\ Phys.\  B {\bf 823}, 174 (2009)
  [arXiv:0907.1625 [hep-th]].




\end{thebibliography}
\end{document}